\documentclass[fleqn,usenatbib]{mnras}

\usepackage{newtxtext,newtxmath}

\usepackage[T1]{fontenc}

\DeclareRobustCommand{\VAN}[3]{#2}
\let\VANthebibliography\thebibliography
\def\thebibliography{\DeclareRobustCommand{\VAN}[3]{##3}\VANthebibliography}

\usepackage{graphicx}	
\usepackage{amsmath}	

\newcommand{\TESSILATOR}{{\sc TESSILATOR}}

\title[\TESSILATOR]{\TESSILATOR: a one-stop shop for measuring TESS rotation periods}

\author[A. S. Binks \& H. M. G\"unther]{
A. S. Binks$^{1,2}$\thanks{E-mail: binks@astro.uni-tuebingen.de (ASB)} $\&$
H. M. G\"unther$^{1}$
\\
$^{1}$MIT Kavli Institute for Astrophysics and Space Research, Massachusetts Institute of
Technology, Cambridge, MA 02139, USA\\
$^{2}$Institut f\"ur Astronomie und Astrophysik, Eberhard-Karls Universit\"at T\"ubingen, Sand 1, 72076 T\"ubingen, Germany
}

\date{Accepted XXX. Received YYY; in original form ZZZ}

\pubyear{2024}

\begin{document}
\label{firstpage}
\pagerange{\pageref{firstpage}--\pageref{lastpage}}
\maketitle

\begin{abstract}

We present a software package designed to produce photometric lightcurves and measure rotation periods from full-frame images taken by the Transiting Exoplanet Survey Satellite (TESS), which we name ``\TESSILATOR''. \TESSILATOR~is the only publicly-available code that will run a full lightcurve and rotation period ($P_{\rm rot}$) analysis based on just a (list of) target identifier(s) or sky position(s) via a simple command-line prompt. This paper sets out to introduce the rationale for developing \TESSILATOR, and then describes the methods, considerations and assumptions for: extracting photometry; dealing with potential contamination; accounting for natural and instrumental systematic effects; lightcurve normalisation and detrending; removing outliers and unreliable data; and finally, measuring the $P_{\rm rot}$ value and several periodogram attributes. Our methods have been tuned specifically to optimise TESS lightcurves and are independent from the pipelines developed by the TESS Science Processing Operations Center, meaning \TESSILATOR~can, in principle, analyse {\it any} target across the entire celestial sphere. 

We compare \TESSILATOR~$P_{\rm rot}$ measurements with TESS-SPOC-derived lightcurves of 1,560 (mainly FGKM-type) stars across four benchmark open clusters (Pisces–Eridanus, the Pleiades, the Hyades and Praesepe) and a sample of nearby field M-dwarfs. From a vetted subsample of 864 targets we find an excellent return of $P_{\rm rot}$ matches for the first 3 open clusters ($>85$ per cent) and a moderate ($\sim 60$ per cent) match for the 700\,Myr Praesepe and MEarth sample, which validates \TESSILATOR~as a tool for measuring $P_{\rm rot}$. The \TESSILATOR~code is available at \url{https://github.com/alexbinks/tessilator}.
\end{abstract}




\section{Introduction}\label{sec:introduction}

All stars (and celestial objects) rotate. The rate at which stars spin is governed by the angular momentum transfer processes that occur as mass density profiles and energy content are redistributed as they evolve. The time taken for a star to complete one full revolution on its rotational axis, commonly referred to as the rotation period ($P_{\rm rot}$), is an age and mass-dependent quantity that is vital to determine the stellar radius and rotational velocity. Rotation is therefore a crucial feature that must be correctly accounted for in evolutionary models. Our lack of understanding of stellar rotation at young ages may be partially responsible for the large discrepancies amongst various models in terms of their age and mass calculations of pre-main sequence (PMS, $5-100$\,Myr) low-mass (spectral types FGK and M) stars \citep{2015a_Bell,2022a_Binks}. Whilst rotation has been known for decades to scale with other features of PMS evolution, such as magnetic activity \citep{1972a_Skumanich,1984a_Noyes,2008a_Mamajek,2021a_Basri} and Lithium depletion \citep{1993a_Soderblom,2021a_Jeffries}, a full picture of the effects that rotation has on these process is yet to emerge. Additional gaps in our understanding have been exposed at older ages, where stars of similar spectral-type are observed to spin-down slower than standard predictions \citep{2019a_Curtis,2020a_Spada}. Whilst a magnetic wind-breaking model has been effective in partially reconciling the discrepancy \citep{2016a_Van_Saders}, the exact physical process that drive the wind-breaking are poorly understood.

Rapid rotation can act as an important indicator of stellar youth for solar-type and low-mass stars, where alternative age-dating methods may be less effective \citep{2015a_Binks}. By measuring $P_{\rm rot}$ and analysing photometric lightcurves, features or properties that are often difficult to detect, such as magnetic activity, flares, (unresolved) multiplicity and age can be placed on a quantitative basis. Models incorporating the evolution of stellar angular momentum \citep{1962a_Schatzman,1967a_Kraft,1988a_Kawaler,1992a_Zahn,2013a_Gallet,2018a_Garraffo,2019a_Amard} predict three main phases of spin evolution before they slowly brake in a quasi-static fashion upon reaching the zero-age main sequence (ZAMS). Stars are born with an initial $P_{\rm rot}$, which is thought to be largely regulated at early ages through the locking of magnetospheric accretion funnels to their protoplanetary disks \citep{1991a_Koenigl,2007a_Bouvier}. Subsequent to disk dissipation \citep[typically after $\sim$10 Myr, through accretion processes and/or UV photoionisation,][]{2009a_Gorti,2011a_Owen,2019a_Picogna} they spin up during PMS contraction. The peak angular velocity and time at which this occurs is mostly mass dependent, where low-mass stars tend to rotate faster, but take longer to spin up. Once Solar-type stars arrive on the ZAMS, angular momentum is gradually lost through mass-loss in stellar winds \citep{1967a_Kraft,1972a_Skumanich}, whereas low-mass stars converge onto the slow-rotating sequence slower and they stay fast-rotating longer \citep{2003a_Barnes}.

Efforts to characterise the distribution of $P_{\rm rot}$ amongst members in many star clusters spanning ages from the early PMS to a few Gyr have verified model predictions to some extent, and provided parameterised empirical models connecting $P_{\rm rot}$, photometric colour and age. This framework, known as ``Gyrochronology'' \citep{2003a_Barnes,2007a_Barnes,2015a_Angus,2019a_Angus}, has the power not only to offer potentially precise ages for clusters (less so for individual stars), but acts as a rigorous test for angular momentum models. There are unilateral findings that $P_{\rm rot}$ scales with: the inverse of X-ray luminosity \citep{1981a_Pallavicini} -- a proxy for coronal activity; $\log R'_{HK}$ or the Ca {\sc ii} triplet -- a proxy for chromospheric activity \citep{1978a_Wilson, 1979a_Linsky, 2018a_Suarez-Mascareno, 2019a_Toledo-Padron}; and the occurence rate of stellar flares \citep{2016a_Davenport, 2020a_Guenther, 2020a_Raetz}.

Up until the early 2000s, the yields of $P_{\rm rot}$ measurements from most surveys were typically on the order of several tens or hundreds, and usually focused on particular regions of sky (e.g., star clusters, OB associations, moving groups). The acquisition of photometric lightcurves can be time consuming, requiring high-cadence time-series data, with observing baselines often lasting months. Whilst spectroscopic rotational velocity ($v\sin i$) measurements were more relatively abundant \citep{2000a_Glebocki, 2014a_De_Medeiros}, the ambiguity of their inclination angles can only provide lower-limits on inferred $P_{\rm rot}$ values. In the past two decades, the number of $P_{\rm rot}$ measurements swelled by several orders of magnitude, partially from dedicated surveys of open clusters \citep[e.g., the MONITOR project,][]{2007a_Irwin, 2008a_Irwin}, but mostly from the analysis of vast quantities of stellar lightcurves from missions whose primary goals focused on detecting exoplanet transits. These include ground-based missions such the Hungarian Automated Telescope \citep[HATNet,][]{2002a_Bakos} and the Super-Wide Angle Search for Planets \citep[SuperWASP,][]{2006a_Pollacco} and satellite surveys such as the Co-Rotating Telescope \citep[CoRoT,][]{2009a_Auvergne}, the latter of which was a pre-cursor to the ground-breaking Kepler (and subsequent K2) (\citealt{2014a_Howell,2016a_Borucki}) and the Transiting Exoplanet Survey Satellite \citep[TESS,][]{2015a_Ricker} missions. This paper focuses on data obtained exclusively from TESS\footnote{A set of targets with Kepler K2 $P_{\rm rot}$ measurements is used in $\S$\ref{sec:R23_comparison} as a comparative sample.}.

Launched in 2018 (and still acquiring data), TESS has so far identified in excess of $5\,000$ planets, and another $\sim$7\,000 are predicted to be discovered \citep{2022a_Kunimoto}. It has collected time-series data for over 95 per cent of the sky at least once (with most regions having $\geq$2 scans) and should be virtually all-sky by late-2024. TESS provides data in two modes during each of its $\sim$27 day observing sectors: firstly the 2-minute cadence lightcurves generated by the TESS Science Processing Operations Center \citep{2016a_Jenkins,2020a_Caldwell} specialised for $\sim2\times10^{5}$ stars selected as prime candidate exoplanet hosts, and secondly the longer cadence full frame image calibrations (FFICs) recorded across the whole detector in the 4 CCDs across 4 cameras that cover a $24^{\circ}\times96^{\circ}$ sky segment. The latter of these TESS modes provides an unprecedented unique opportunity to acquire lightcurves for stars across the full celestial sphere.



Many methods have been developed to measure stellar $P_{\rm rot}$, such as the generalised Lomb-Scargle periodogram \citep[LSP,][]{1976a_Lomb,1982a_Scargle,2009a_Zechmeister,2018a_VanderPlas}, Autocorrelation Function \citep[ACF,][]{2013a_McQuillan,2022a_Holcomb}, Wavelet Transforms \citep[WTs,][]{2009a_Carter,2014a_Garcia,2022a_Claytor}, Gaussian Processes \citep[GPs,][]{2018a_Angus,2023a_Aigrain} or a mixture of these \citep{2017a_Ceillier,2019a_Santos,2020a_Reinhold,2024a_Colman}. Each of the methods have strengths and weaknesses. For the purpose of this work, we focus only on the LSP method.

Given the potential for TESS to provide $P_{\rm rot}$ values for an enormous number of stars, and the impact they could have on improving our understanding of stellar rotational evolution, we have developed a software package named ``\TESSILATOR''. It is specifically designed as an all-in-one program which, for any given target (or list of targets) will: automatically download the TESS data; quantify contamination from neighbouring sources; perform aperture photometry and generate lightcurves; normalise, detrend and clean lightcurves; correct for general systematic features (if required); perform a LSP analysis to measure the $P_{\rm rot}$ and several parameters to assess the quality and reliability of the measurements, and evaluate a final $P_{\rm rot}$ for targets with multiple TESS sectors. In this paper we present \TESSILATOR~to the community. We provide a detailed description of \TESSILATOR's functions, the lightcurve and periodogram analysis and test the performance of the code by comparing $P_{\rm rot}$ measurements for a set of open clusters and nearby M-dwarf field stars.

In $\S$\ref{sec:rationale} we briefly outline the rationale for developing bespoke software for analysing TESS lightcurves. In $\S$\ref{sec:tessilator_code}~we describe how \TESSILATOR~performs each step described in the previous paragraph, and the underlying assumptions and considerations that lead to a processed lightcurve and $P_{\rm rot}$ measurement. To show how the \TESSILATOR~performs in practice, in $\S$\ref{sec:period_comparison}~we compare \TESSILATOR~$P_{\rm rot}$ measurements for 1560 stars with results from a recent publication and also a large sample of field stars from several recently published variability surveys. We provide our conclusions and plans for future \TESSILATOR~projects in $\S$\ref{sec:conclusions}.

\section{Software for analysing TESS data}\label{sec:rationale}

The primary objective of \TESSILATOR~is to: perform aperture photometry for TESS image data; apply cleaning and detrending procedures to improve lightcurves; measure the $P_{\rm rot}$ using LSP analysis; and assess the quality and reliability of these results by accounting for the background contamination and various properties of the lightcurve and periodogram output. All this can be done with a single command line prompt, where the user is required only to provide the target name, or a text file containing a list of target names (or sky-positions, see $\S\ref{sec:target_input}$).

\subsection{Pre-existing TESS-based public codes}\label{sec:other_codes}
Whilst many software tools have been developed for analysing TESS lightcurves from FFIC data, here we simply highlight some of the most frequently used software packages and discuss their relative merits. Two of the most commonly cited TESS codes are ``eleanor''\footnote{\url{https://adina.feinste.in/eleanor/}} \citep{2019a_Feinstein} and ``Lightkurve''\footnote{\url{https://docs.lightkurve.org/index.html}} \citep{2018a_Lightkurve}. Whilst both these pipelines perform a careful lightcurve extraction by correcting for the instrument systematics apriori and applying sophisticated point-spread function (PSF) photometry, the former is optimised for planet searches (in line with the main aim of the TESS mission) and does not measure $P_{\rm rot}$ and the latter, whilst offering a periodogram function to measure $P_{\rm rot}$, does not provide any quality or reliability metrics for the lightcurve or periodogram. Neither of these tools provide means to measure $P_{\rm rot}$ directly from receiving simply a target as input, requiring the user to incorporate functions into their own code. Moreover, neither of these codes account for the effects of neighbouring sources which can strongly contaminate the measured flux for a given target. A similar code, {\sc tessextractor}\footnote{\url{https://www.tessextractor.app/}} \citep{2021a_Serna} is also available, but uses different methods to measure $P_{\rm rot}$ and does not provide a detailed output of the lightcurve vetting and periodogram results.

Two additional TESS codes have been recently published. The first one, developed by \citet[][herein, R23]{2023a_Rampalli}, use TESS lightcurves that have been pre-vetted by TESS-SPOC and directly perform LSP to obtain $P_{\rm rot}$. Their method includes a training algorithm that utilises Kepler-K2 data to reduce false positives and true negatives. We adopt similar vetting processes used by R23 to determine a final $P_{\rm rot}$ measurement in the case of multiple sectors and in $\S$\ref{sec:R23_comparison}~we run \TESSILATOR~on the R23 sample and directly compare $P_{\rm rot}$ values. The second recent TESS code, {\sc spinneret} \citep{2024a_Colman} uses exclusively 2-minute cadence TESS data and measures $P_{\rm rot}$ using 2 different types of periodogram analysis and the autocorrelation method (ACF). They employ a sophisticated random forest classifier algorithm to quantify a subset of several thousand reliable periods. Similar methodologies are being considered for future \TESSILATOR~releases.

Since the length of a TESS sector is only $\sim 27\,$d, and that systematic offsets are introduced when data is downlinked from the satellite, TESS struggles to measure $P_{\rm rot}$ for targets with $P_{\rm rot} > 10\,$d, especially for faint lightcurves. Combining data from multiple sectors is generally not possible because the instrumental response between sectors is generally non-uniform (with the exception perhaps of the TESS Continuous Viewing Zone, \citealt{2022a_Hattori}). Despite these limitations, specialised tools have been made in an attempt to improve sensitivity in the $P_{\rm rot} \sim 10-15\,$d range \citep[e.g.,][see \citealt{2024a_Colman} for a detailed summary of these tools]{2020a_Hedges,2022a_Claytor,2022a_Hattori,2022a_Holcomb}. Comparisons with targets observed in photometric missions with longer observing baselines (e.g., {\it Kepler, MEarth, Zwicky Transient Facility}) show $P_{\rm rot}$ matching rates of up to 10\,per cent. We do not incorporate any of these methods in \TESSILATOR, but recommend readers to try these codes for measuring longer $P_{\rm rot}$ with TESS data.

\subsection{Why do we need \TESSILATOR?}\label{sec:why_tessilator}
Despite the abundance of available TESS codes, \TESSILATOR~fills an important gap for TESS data analysis. It provides a vital resource for measuring $P_{\rm rot}$ values from an extremely rich, all-sky reserve of photometric data. The ease of usage, combined with the capability of analysing large catalogues of targets means \TESSILATOR~can provide important tests for many aspects of our understanding of stellar rotational evolution, and will contribute to tackling those major outstanding problems addressed in the first paragraph of $\S$\ref{sec:introduction}. Whilst \TESSILATOR~can be run directly from the command line the code is entirely comprised of Python functions and all variable parameters are set as keywords that can be modified from their default values, providing users with flexibility to call individual functions into their own code. In summary, \TESSILATOR~has unique features compared to other available codes because of the following reasons:
\begin{itemize}
\item {\bf Ease of use} -- \TESSILATOR~can be easily installed and used directly from the command line.
\item {\bf Large target lists} -- a large survey using \TESSILATOR~analysed $>3$~million lightcurves and measured $P_{\rm rot}$ for $\sim$1.1 million targets. The process took $<2$ weeks using a moderate cluster with 208 cores (Binks et al. 2024, in preparation).
\item {\bf Treatment of contamination} -- a specialised algorithm has been implemented to quantifying contaminating flux from neighbouring sources (see $\S$\ref{sec:contamination}).
\item {\bf Corrections to systematics} -- 2 separate routines have been developed to correct for systematic trends, if required.
\item {\bf Lightcurve vetting} -- multiple functions have been created to carefully modify lightcurves and remove spurious data.
\item {\bf Periodogram analysis} -- as well as performing the LSP and identifying multiple peaks as potential rotational signals, \TESSILATOR~also offers a specialised periodogram shuffling method to detect short periods in noisy lightcurves (see $\S$\ref{sec:shuffled_period}).
\item {\bf Outputs and products} -- \TESSILATOR~is able to store the downloaded fits files, it provides aperture photometry, lightcurve and periodogram data, and produces informative summary plots for the analysis of each target.
\item {\bf Final period selection} -- a specialised routine is offered that selects a final $P_{\rm rot}$ measurement in the case where a target has two or more sectors of TESS data (see $\S$\ref{sec:final_period}).
\end{itemize}

\section{The \TESSILATOR~code}\label{sec:tessilator_code}

The \TESSILATOR~code, version 1.0, is available to download and the software\footnote{\url{https://github.com/alexbinks/tessilator}} and supporting documentation\footnote{\url{https://tessilator.readthedocs.io/en/latest/}} are regularly updated. We anticipate future releases that incorporate additional features, which will be largely based on feedback from users. We refer readers interested in the software architecture to the supporting documentation, which includes an application programming interface (API) with detailed descriptions for each function. 

\begin{figure*}
    \centering
    \includegraphics[width=0.98\textwidth]{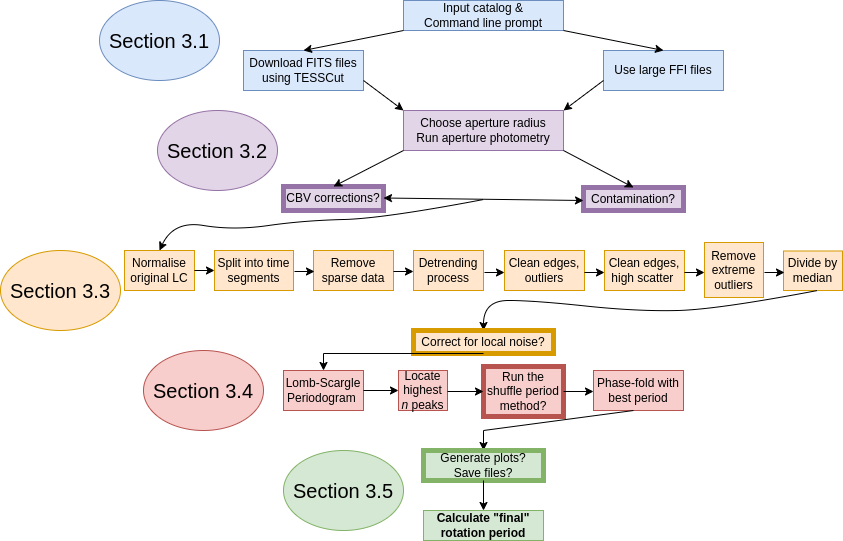}
    \caption{A flowchart depicting the procedures performed by \TESSILATOR. The colours are used to group small procedures into higher-level, general steps in the code, as follows: blue=target input and data retrieval ($\S$\ref{sec:target_input_data_retrieval}); purple=aperture photometry ($\S$\ref{sec:aperture}); orange=lightcurve analysis ($\S$\ref{sec:lightcurve_analysis}); pink=$P_{\rm rot}$ calculation ($\S$\ref{sec:periodogram_analysis}) and green=``final'' $P_{\rm rot}$ for a given target ($\S$\ref{sec:outputs_and_finalprot}). Inputs with bold outlines are optional steps.}
    \label{fig:flowchart}
\end{figure*}

To guide the reader through the processing steps in this section, we present examples from the set 6551 \TESSILATOR~lightcurves for 1560 targets selected for TESS lightcurve analysis in R23. The details of the R23 sample are described in $\S$\ref{sec:R23_comparison}. In this section, we describe each step carried out by \TESSILATOR, from quantifying contamination, performing aperture photometry, normalising, cleaning and detrending lightcurves, removing systematic flux, applying the LSP, returning quality and reliability parameters, and calculating a final $P_{\rm rot}$. In each step, we outline the methods, considerations and underlying assumptions. A flow-chart outlining each step carried out in the \TESSILATOR~code is presented in Figure~\ref{fig:flowchart}.

\subsection{Target input and data retrieval}\label{sec:target_input_data_retrieval}
\subsubsection{Target input}\label{sec:target_input}
The first task is to provide \TESSILATOR~with input, and (if needed) perform a cross-match with the Gaia DR3 main source catalogue \citep[][which is required for assessing the background contamination, see $\S\ref{sec:contamination}$]{2023a_Gaia_Collaboration}.
If only a single target is required, this may be passed as a command-line argument as either a resolvable target identifier or as a pair of sky coordinates. \TESSILATOR~can read file input in three different modes: (a) target identifiers; (b) sky coordinates; or (c) a list containing basic Gaia data.

\subsubsection{Data retrieval}\label{sec:data_retrieval}
The code extracts TESS full-frame images (calibrated, FFIC) 
in two separate modes: either by downloading small regions of data surrounding each target of interest by using the {\sc TESSCut} software \citep[][herein, the cutout method]{2019a_Brasseur}, or by using the entire set of FFIC images for a given Sector/Camera/CCD (SCC) configuration\footnote{Available for download at \url{https://archive.stsci.edu/tess/bulk_downloads/bulk_downloads_ffi-tp-lc-dv.html}} and locating positions of target stars on these images using the {\sc tess\_stars2px.py} module from \citet[][herein, the sector method]{2020a_Burke}. If the cutout method is chosen, the user can decide whether to collect data for all available sectors, a specified list of sectors, or simply just one sector, and also has an option to cap the number of sectors for download (this is useful when there are many TESS sectors for a given target or group of targets).

The observing cadence for the FFICs was shortened from $\sim$30\,min between sectors 1-27 (cycles 1 \& 2, the primary mission), to 10\,min between sectors 28-55 (cycles 3 \& 4, the first extended mission) and to 200\,s from sectors 56 onwards (cycles 5+, the second extended mission). For the sectors with 30\,min cadence, each {\sc TESSCut} image, centered on a 21x21 pixel tile requires $\sim$9\,MB of disk space (the file size is inversely proportional with cadence), whereas the sector method for a given SCC requires $\sim$36\,GB of disk space. Therefore for surveys with $\gtrsim 4000$ targets in a given SCC, the sector method is better in terms of memory storage. However, even for projects with slightly fewer targets one may still want to consider the sector method because of the time expense of downloading many individual sources.

Ultimately, the preferred choice of data extraction depends on the task in hand, but as a general guide, smaller or larger surveys are better suited to the cutout or sector method, respectively. The reader should note that a full sector of FFICs requires 16 times more disk space than a single SCC, because there are 4 cameras and 4 CCDs per sector. Once data has been downloaded and ingested by \TESSILATOR, the subsequent processes are identical for each of these two modes.

\subsection{Aperture photometry}\label{sec:aperture}

Since the aim of \TESSILATOR~is primarily to analyse lightcurves for (faint) stars, we treat targets as point sources and use circular aperture photometry to calculate instrumental TESS $T$-band fluxes for all individual frames. Depending the choice of FFIC data extraction, these could be from the image stack produced by {\sc TESSCut} or a list of time-ordered images from a full SCC set. We tested and confirmed that both methods would produce identical lightcurves.

Using a $21\times21$ pixel region with the aperture centered on the pixel closest to the given sky-position, an annulus region with inner and outer radii of 6 and 8 pixels (respectively) is used to calculate the median background flux per pixel. The chosen annuli values ensure the background contains minimal flux contribution from the source, whilst the annulus area is large enough that any bright background sources would not affect the median count.

\begin{figure*}
    \centering
    \includegraphics[width=0.98\textwidth]{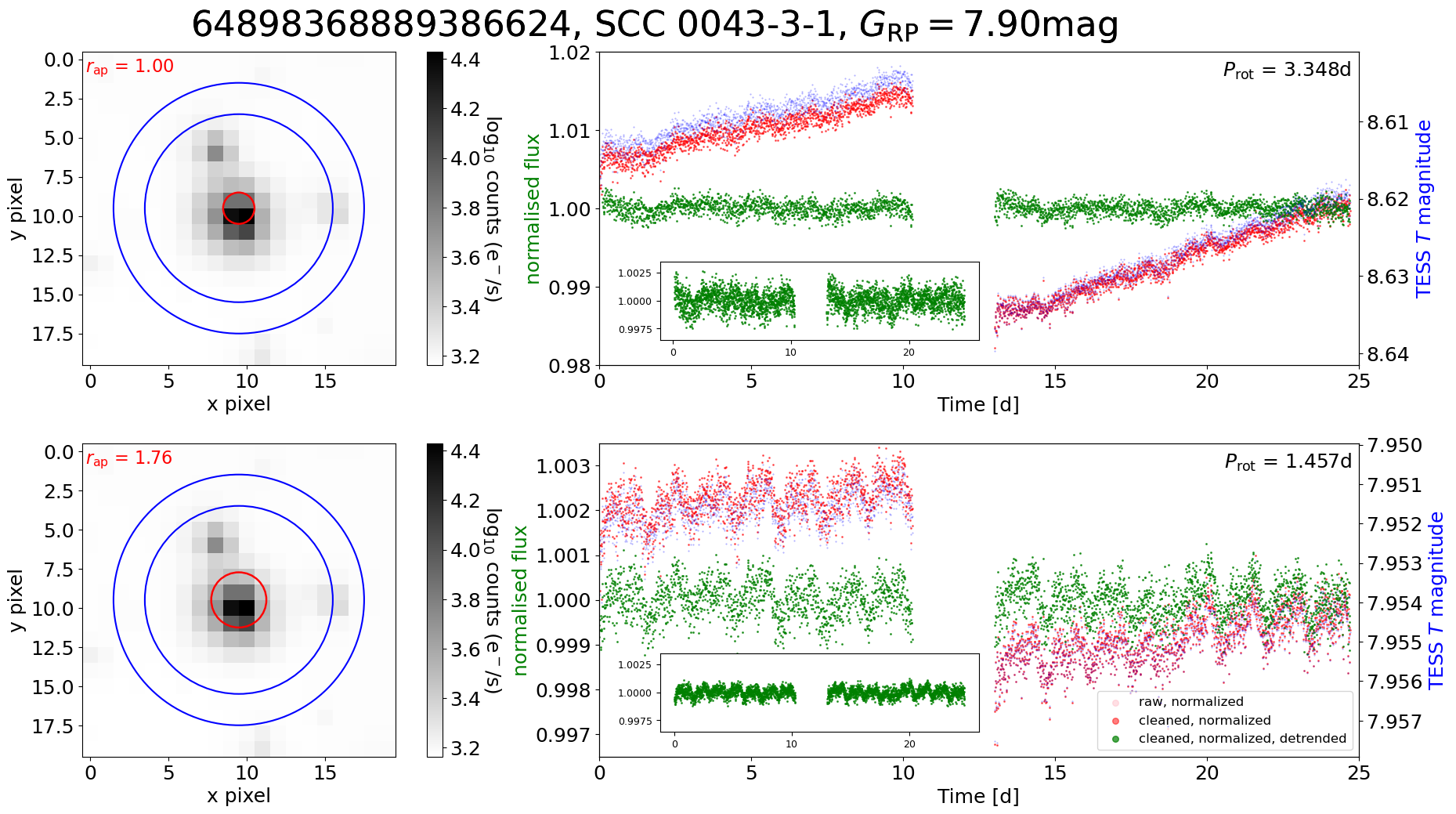}
    \caption{An example depicting how the choice of aperture radius can affect the (normalised, detrended and cleaned -- green data points, see $\S$\ref{sec:lightcurve_analysis}) lightcurve and $P_{\rm rot}$ calculation for Gaia DR3 64898368889386624 (Sector 43, Camera 3, CCD 1), a bright target with $G=8.124$\,mag. The left panels show the 20x20 pixel cut-out image centered on the target, with the aperture and annuli used to evaluate the source and background flux are shown as one red and two blue circles, respectively. The right panels represent the TESS magnitude, normalised flux, and the subsequent detrended flux (see $\S$\ref{sec:lightcurve_analysis}) as blue, red and green points, where the insets are both zoomed-in regions using the same range of y-values to indicate the amplitude of variability for the detrended flux. The results for a (fixed) aperture radius of 1.0 pixel, and that of a 1.76 pixel extraction (based on the algorithm described in $\S$\ref{sec:aperture}) are displayed in the top and bottom panels, respectively. The larger aperture radius is clearly preferential, not only because the measured TESS $T$ magnitude is better matched to the Gaia DR3 $G_{\rm RP}$ magnitude ($=7.900$\,mag, the TESS magnitude filters are closely matched with Gaia DR3 $G_{\rm RP}$, see $\S$\ref{sec:contamination}), but the measured $P_{\rm rot}$ ($=1.457\,$d) is in good agreement with the R23 value ($=1.462\,$d), whereas the top panel fails to provide a match to either of these parameters.}
    \label{fig:aperture_radius}
\end{figure*}

There are two options for selecting the aperture radius ($r_{\rm ap}$). To calculate the amount of flux incident within the aperture, we measure the fractional area of each pixel included in the aperture, and take this fraction of the flux from each contributing pixel. The first option is to use a fixed value, which is set as default to 1.0 pixel. This, incidentally, is close to the theoretical optimum for flux extraction in the sky-limited domain \citep{1998a_Naylor}. The second option uses a simple algorithm we developed called ``calc\_rad'', which selects $r_{\rm ap}$ by assessing whether the immediate neighbouring pixels (i.e., the 1-pixel border surrounding a square) contribute significant flux compared to the central pixel. This is often the case for bright stars, where $r_{\rm ap}=1$ would not capture enough of the source flux, and can lead to erroneous lightcurve production (see, e.g., Figure\,\ref{fig:aperture_radius}).

Here we describe in basic terms how ``calc\_rad'' calculates $r_{\rm ap}$. The flux values are initially background subtracted 
and a variable called $r$ is set equal to 1 (representing the start of the procedure). If $k_{r+1}$, the ratio of the median value of the 8 neighbouring pixels and the central pixel is greater than some specified value, $k_{\rm lim}$ (the default value is $0.1$), $r$ increases by 1, and the flux values of the next set of 16 surrounding pixels are compared with the central pixel, and so on, until $k_{r+1}<k_{\rm lim}$. The final value for $r_{\rm ap}$ is calculated by linearly interpolating between $k_{r}$ and $k_{r+1}$, using the $r$ and $r+1$ values either side of the pass/fail boundary and finding the value for $r$ where $k_{r}=k_{\rm lim}$. If, after 3 iterations, the $k_{r}>k_{\rm lim}$ condition is still satisfied, or if $k_{r}$ increases at any point, then we suspect there might be contribution from a bright neighbouring source and we reset the final $r_{\rm ap}$ to 1 pixel. This algorithm is repeated for every image in the stacked data frame. Altering $r_{\rm ap}$ between frames would cause discontinuities in the lightcurve, therefore the median aperture radius is chosen as the final $r_{\rm ap}$ to be used in each image frame.

Once $r_{\rm ap}$ has been evaluated, aperture photometry is performed using methods from the {\sc photutils} Python package \citep{2022a_Bradley}. The background flux per pixel is multiplied by the aperture area to calculate the total background flux in the aperture, and $T$ magnitudes are calculated using a zero-point of $20.44\pm 0.05$\,mag \citep{2018a_Vanderspek}. Only images with data quality flags (fits header: ``DQUALITY'') set to zero are retained in the analysis. Once aperture photometry has been performed for all images, the results are stored to a table ready for further vetting. It is important to point out that when using the sector version of \TESSILATOR, aperture photometry can be simultaneously performed at multiple positions on the FFICs, which significantly expedites the image processing.

\subsubsection{Background contamination}\label{sec:contamination}

Due to the large size of TESS pixels (the pixel length corresponds to an angular extent of 21''), contamination from neighbouring sources can be an important factor that reduces the signal to noise for a target source and/or can result in a false $P_{\rm rot}$ measurement for a target, where the true source of the $P_{\rm rot}$ is from a neighbouring contaminant.

To quantify the flux contribution from neighbouring sources that passes through $r_{\rm ap}$ (defined in $\S$\ref{sec:aperture}), \TESSILATOR~performs an SQL search of the Gaia DR3 catalogue to find neighbouring sources within a given pixel distance of the target. Since the passbands and response functions for TESS $T$ \citep{2015a_Ricker} and Gaia DR3 $G_{\rm RP}$ \citep{2021a_Riello} photometry are very similar ($600-1000\,$nm), and given that the practical faint limit for measuring TESS lightcurves is much brighter than the faint limit of $G_{\rm RP}$ in Gaia DR3, we use $G_{\rm RP}$ fluxes to represent the source contamination.

\begin{figure}
    \centering
    \includegraphics[width=0.48\textwidth]{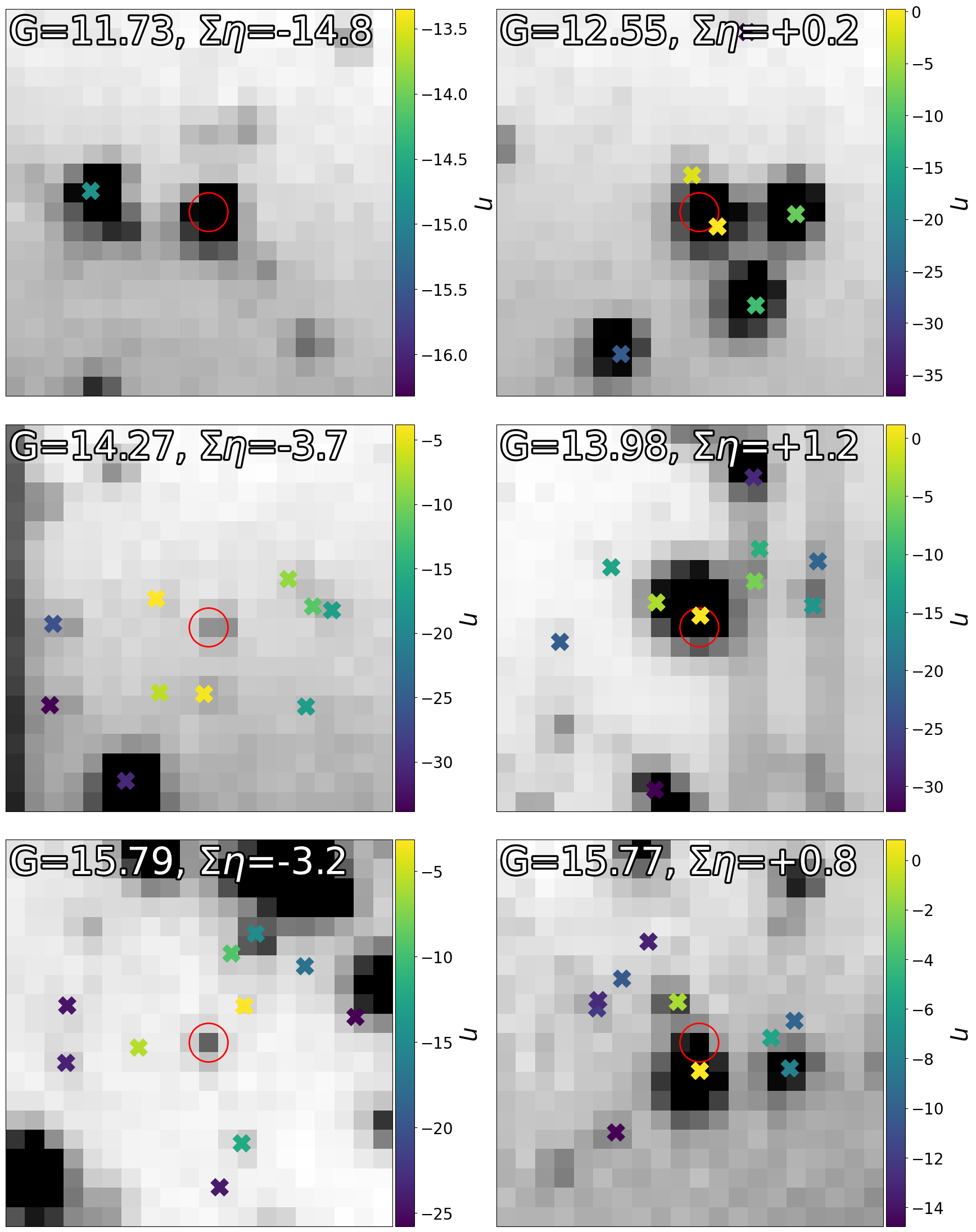}
    \caption{FFIC images for 6 targets, indicating contamination from neighbouring sources. The top, middle and lower panels represent bright ($11<G<13$), medium ($13<G<15$) and faint ($G>15$) targets, and the left and right panels show examples of low ($\Sigma\eta < 0$) and high contamination ($\Sigma\eta > 0$), respectively. The coloured crosses on each image represent the X/Y pixel positions of the top 10 contributing contaminants, in terms of $\eta$ (see $\S$\ref{sec:contamination}).}
    \label{fig:contamination}
\end{figure}

The procedure is identical to \cite{2022a_Binks}, who incorporate equation 3b-10 of \cite{1965a_Biser} to calculate the amount of incident flux within a given aperture from a source at a given distance. The only exception in \TESSILATOR's case is the use of a different Gaia filter (Gaia DR2 $G$-band fluxes are used in \citealt{2022a_Binks}). For practical purposes we set a faint limit for neighbouring sources at 3 magnitudes fainter than the target, and set a search radius of 10 pixels (as a default value). The program returns the (base-10 logarithm of the) flux ratio between the sum of the neighbouring sources and the target ($\Sigma\eta$) and also the brightest contaminant and the target ($\eta_{\rm max}$). Examples of sources with low and high contamination are presented in Figure~\ref{fig:contamination}. Each contaminant is ordered in descending values of $\eta$ and the first few of these are stored to a file (the default is 10).

An optional procedure is offered that assesses whether any of these stored contaminants might be the source of the measured $P_{\rm rot}$ and/or if the target $P_{\rm rot}$ might be deemed unreliable. Lightcurve processing and periodogram analyses are performed for each contaminant, using the exact same steps ($\S\ref{sec:lightcurve_analysis}$ and $\S\ref{sec:periodogram_analysis}$) as those made for the target. Each contaminant is assigned a ``false flag'' and a ``reliability flag''. A false flag value of 1 (or 0) is assigned if the absolute difference between the contaminant and target $P_{\rm rot}$ is less (more) than the quadrature sum of the $P_{\rm rot}$ uncertainties. The reliability flag depends on the pixel distance ($d_{\rm pix}$) between the contaminant and target. If $d_{\rm pix} < 1$, the contaminant is considered unresolved, so if the contaminant flux is greater (less) than the target flux, we deem this as unreliable (reliable) and assign a value of 1 (0). In the case where $d_{\rm pix}>1$, if the amplitude of the phase-folded lightcurve of the contaminant is greater (less) than half that of the original target lightcurve, we assign a value of 1 (0). If an error occurs during these tests, or if there are no contaminants, or if this option is ignored, then both flags return a value of 2, 3 or 4, respectively. These flags can be particularly useful if the target is in a crowded field or if the $P_{\rm rot}$ reliability is particularly important. However, the additional computation time needed scales with the number of contaminants to be assessed, and the processing of many targets in this way can become cumbersome.

\subsubsection{Correcting for systematics: Cotrending Basis Vectors}\label{sec:CBV_corrections}

During image exposures, TESS CCDs are sensitive to radiation from sources other than the targeted sky region. These contribute to a systematic flux component. Some of this systematic flux comprises of scattered diurnal and lunar light, the former of which can impart small 1-day aliasing in the lightcurves, whereas the latter contributes an additional noise term. The other systematic effect comes from the instrument response, which causes non-uniform flux sensitivity across the CCD chip. This is often due to temperature gradients across the light collecting area, especially at times just before or after data transfer when the detectors are switched off and back on. In addition, the CCD chips suffer degradation over time. Systematic flux is likely to be more problematic for targets with low signal-to-noise ratio or for targets with very low amplitudes of variability.

The TESS science team have provided tools to help reduce the systematic flux imprint. By measuring the lightcurves for many stars across each SCC, common features in the lightcurves are detected using principal component analysis. These are provided as fits tables each containing 8 co-trending basis vectors (CBVs) sampled at shorter cadences than the FFI image\footnote{\url{https://archive.stsci.edu/tess/bulk_downloads/bulk_downloads_cbv.html}}. By applying linear fits between the CBVs and the lightcurve, the systematic flux can be removed. The CBV corrections are particularly suited to bright stars and targets whose point spread function is non-circular, and often represent an important stage in producing cleaned lightcurves for transiting planets \citep{2021a_Guerrero} and short-term variable targets including Type Ia supernovae \citep{2021a_Fausnaugh} and blazars \citep{2024a_Poore}. In practice, however, this technique is highly prone to overfitting, and the net effect of CBV fitting can be unwanted noise injection, or the inclusion of lightcurve features that were not initially present.

Therefore, we offer CBV fitting as an option in \TESSILATOR, but advise that the user only employs CBV corrections if they are confident that the effects will improve lightcurve quality. To make the CBV corrections to the lightcurve we apply the method used by \cite{2017a_Aigrain}, which uses a variational Bayes method to optimize the number of CBVs used in the linear fitting procedure. This maximum-likelihood method, designed for Kepler data but applicable to TESS lightcurves, is designed to be robust to overfitting and artificial noise injection.

If a user selects the CBV-correction procedure, the CBV-corrected lightcurve must pass two separate tests for it to be selected as the data to be used to calculate $P_{\rm rot}$, otherwise the original lightcurve is used and this CBV-corrected lightcurve is rejected (although the CBV-corrected lightcurve can be saved to file). The first test (CBV test 1) involves comparing different features of each lightcurve before they have been passed to any of the processing steps described in $\S$\ref{sec:lightcurve_analysis}, and the second test (CBV test 2) uses features from the LSP analysis, where the inputted lightcurves have been fully processed (described in $\S$\ref{sec:periodogram_analysis}). Both tests use a point scoring system based on several criteria. The CBV-corrected lightcurve must score higher than the original lightcurve in CBV test 1 for it to qualify for CBV test 2. The two tests are described in more detail in Appendix~\ref{sec:CBV_tests}, since the reader will need to read sections $\S$\ref{sec:lightcurve_analysis}~and $\S$\ref{sec:periodogram_analysis}~to understand the specific details of each test.

\begin{figure*}
    \centering
    \includegraphics[width=0.9\textwidth]{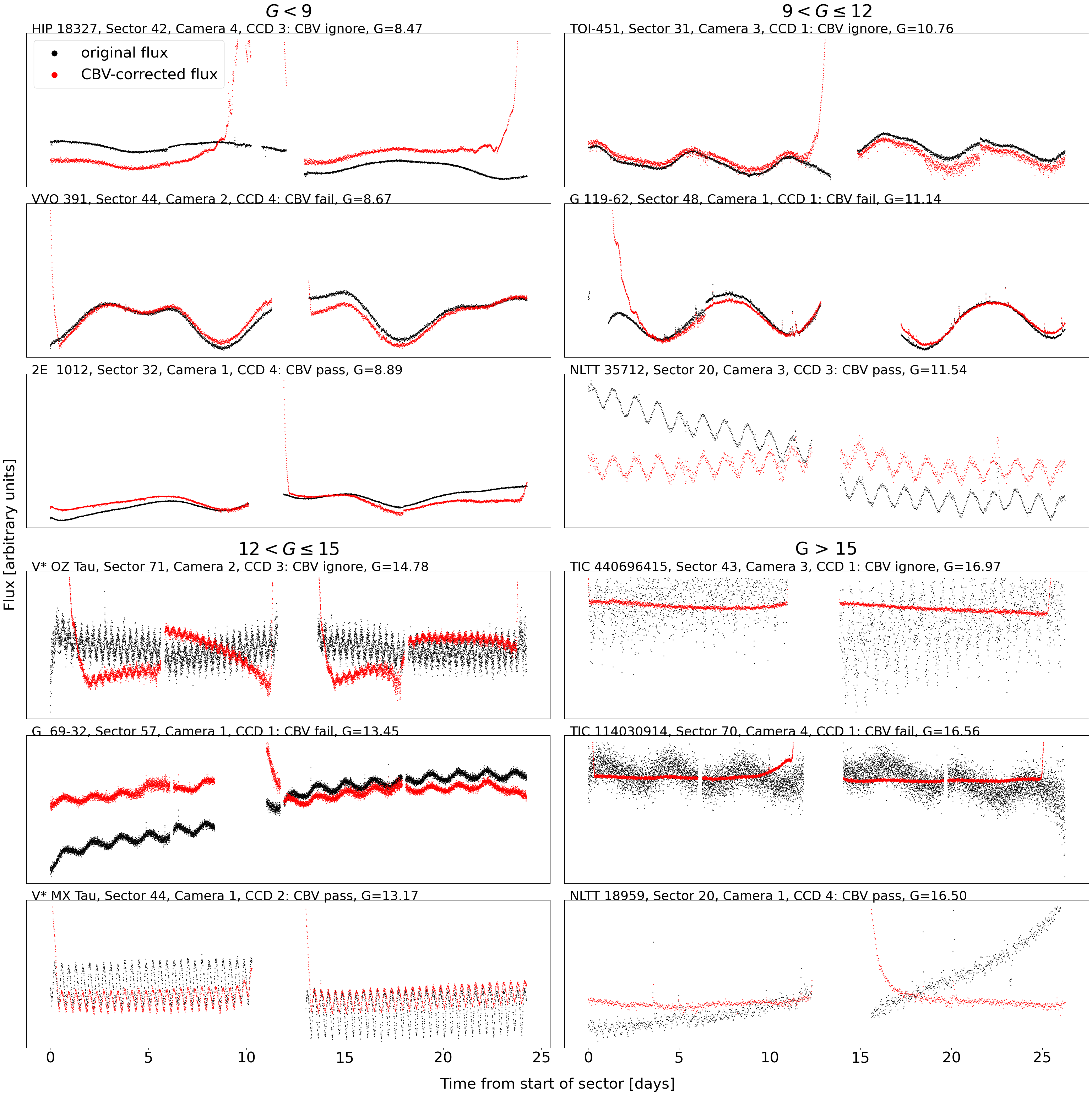}
    \caption{Examples of lightcurve of targets selected from R23, in the cases where the CBV-corrected lightcurve: fails CBV test 1 (``CBV ignore'', first and fourth row), passes CBV test 1, but fails CBV test 2 (``CBV fail'', second and fifth row); or passes both tests (``CBV pass'', third and sixth row). The data from the original fluxes and the CBV-corrected fluxes are highlighted in black and red, respectively. The ``CBV ignore'' example in the $G<9$ set typifies an example of severe over-fitting, where clearly the CBV-corrections result in worse lightcurves. The third and sixth row show that even when both ttests pass, the CBV corrected ligthcurves sometimes have unphysical spikes near the edges that are not present in the original lightcurves.}
    \label{fig:cbv}
\end{figure*}

The CBV-corrected lightcurves therefore fall into three categories, which are labelled and stored as a ``CBV\_choice'' flag. These are: CBV\_choice=0 (rejected in CBV test 1); CBV\_choice=1 (fails CBV test 2); or CBV\_choice=2 (passes both tests). If the CBV-correction option is not selected, then CBV\_choice=3. Examples of comparisons between the original and CBV-corrected lightcurves are provided in Figure~\ref{fig:cbv}. It is clear there are cases where the CBV-corrections serve only to inject noise and create more systematic offsets into the lightcurve (e.g., top panel), cases where the CBV-corrections improve the shape of the lightcurve, but the injection of noise causes the original lightcurve to perform better in the periodogram test (middle panel), and finally cases where the CBV-corrections provide better results (bottom panel).

\begin{figure}
    \centering
    \includegraphics[width=0.48\textwidth]{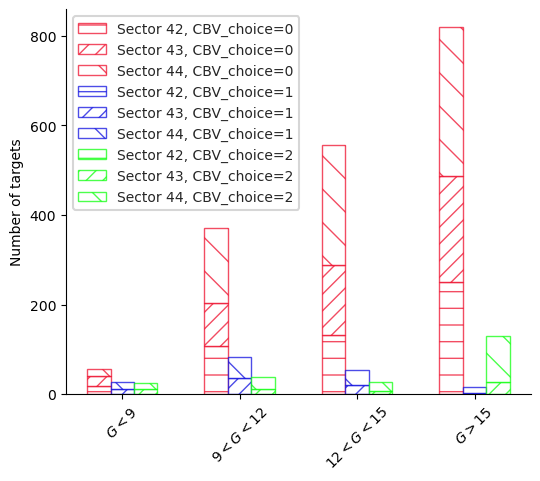}
    \caption{A bar chart representing the proportion of Pleiades targets from R23 that have CBV\_choice labels of either 0 (fail CBV test 1), 1 (fail CBV test 2) or 2 (passes both tests). These results are separated into 4 Gaia DR3 $G$ magnitudes ranges and 3 TESS sectors (42, 43 and 44). The brightest sample ($G<9$) contains the largest fraction of lightcurves labelled as ``CBV pass'' ($\sim 23$\,per cent), whereas all other samples have much lower fractions (all $\lesssim 10$\,per cent). These numbers are fairly constant in each sector, suggesting that the CBV tests do not depend on the sector.}
    \label{fig:cbv_stats}
\end{figure}

To examine the typical outcome of these CBV tests, we use the Pleiades targets from R23 (sectors 42--44). Figure~\ref{fig:cbv_stats}~indicates that, from a possible 2203 lightcurves, only 10 per cent (220) pass both CBV tests. Brighter targets (i.e., those with $G<9$) are labelled ``CBV pass'' slightly more frequently than fainter ones ($\sim 23$ versus $\sim 9$ per cent, respectively), suggesting that the typical net effect of CBV corrections for fainter sources is to inject additional noise. Therefore, we reiterate that CBV corrections should only be applied if the user is confident that they result in a significant improvement to the lightcurve.

\subsection{Lightcurve analysis}\label{sec:lightcurve_analysis}

\begin{figure*}
    \centering
    \includegraphics[width=0.9\textwidth]{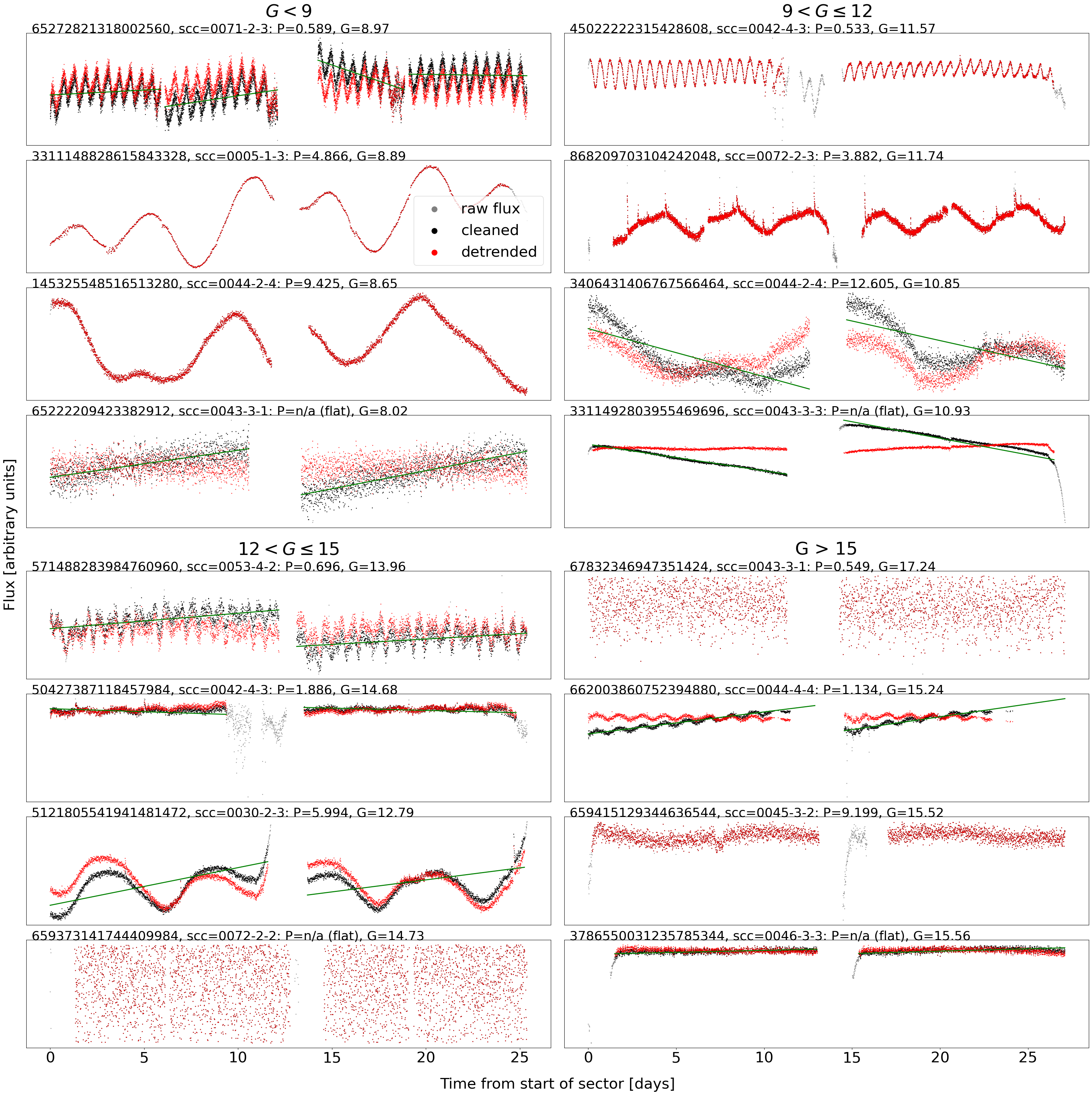}
    \caption{A gallery of \TESSILATOR~lightcurves for 16 targets, selected to provide visual aids to help explain the processes described for the lightcurve analysis in $\S$\ref{sec:lightcurve_analysis}. The top-left, top-right, bottom-left and bottom right quadrants of the gallery represent 4 brightness regimes (in terms of Gaia DR3 $G$\,mag). For each brightness quadrant, in the first, second and third panels we show a fast, medium and slow rotator (where $P_{\rm rot}$ is $<1\,$d, between $1-5\,$d and $>5\,$d, respectively), and the bottom panel represents a lightcurve with an indeterminate $P_{\rm rot}$. Grey and black dots indicate the normalised flux before any processing (raw flux) and after the cleaning steps. Green lines indicate any detrending functions (if used), and the red symbols represent the data that constitutes the final lightcurve.}
    \label{fig:lightcurves}
\end{figure*}

After performing aperture photometry and generating an initial lightcurve for a target (and applying the CBV-correction if required), \TESSILATOR~performs several normalisation, detrending and cleaning procedures to prepare the target for periodogram analyses. In the following sub-sections, we use examples of lightcurves in Figure~\ref{fig:lightcurves}~as a visual aid to describe the processing steps, which will follow the flowchart presented in Figure~\ref{fig:flowchart}.

\subsubsection{Normalising the original lightcurve}\label{sec:normalisation}

The lightcurves are initially normalised, by dividing all (background-subtracted) flux values by the median flux in each lightcurve. The median was selected to ensure robustness to outliers, whilst ensuring the lightcurve amplitudes are not flattened in the normalisation process. Normalised flux error bars are calculated simply by dividing flux uncertainties by their flux values.

\subsubsection{Splitting data into time segments}\label{sec:time_segments}

Whilst the total time baseline per sector for a TESS sector is approximately 27 days, data acquisition for targets observed takes place across 2 parts, both equalling the length of a 13.7 day orbit. Throughout the primary mission and first extended mission (TESS cycles 1-4), data was downlinked from the satellite, typically in $\sim 1-2$ days, whilst close to perigee. Since the second extended mission began (TESS cycles $5+$) the data acquisition strategy changed slightly, such that data is downlinked every $\sim 7$ days, causing 3 gaps, but each with substantially shorter downtime.

During the downlink, the satellite must also perform additional maintenance and engineering tasks such as momentum dumps to ensure the orbit remains stabilised. Therefore the data gap duration changes from sector to sector. We refer to each group of data observed between these gaps as ``segments''. As the CCDs switch observing modes, electronic signals cause temperature variations across the detectors, and this instrumental response frequently leads to discontinuities at the end points of each segment. In addition, across each observing segment, gradual flux gradients are often seen. Figure~\ref{fig:lightcurves} highlights an example of the features described in the former and the latter, in the second panel of the $9<G<12$ group and the top panel of the $12<G<15$ group, respectively.

In order to vet the lightcurves, and remove systematic trends, we must first group the data by their segments. These are defined as sets of contiguous data points whose neighbouring data points are separated in time by no more than 10 times the median time difference. Using this method almost always returns 2 or 4 segments (depending on the TESS mission when the observations were taken), however there are also cases where additional time gaps exist due to excluding poor quality data in the aperture photometry, or simply gaps due to the satellite failing to record photometry, both of which can result in a different number of segments.

\subsubsection{Removing sparse data}\label{sec:cleaning_sparse_data}

Any segments from lightcurves comprising of 2 or more segments that have very few data points ($N_{\rm seg}$) should be removed, as these are typically very scattered and only contribute noise to the lightcurve. We developed a method to remove the sparse data. Firstly, we require a segment to contain at least a minimum number of data points, $N_{\rm thr}$, which is selected as the maximum value between 5\,per cent of the total number of data points in the lightcurve and 50 (both of which are default values in the code). For lightcurves with $>2$ segments, we calculate the standard deviation of $N_{\rm seg}$ values, and if this is larger than $N_{\rm thr}$, we retain only the segments where $N_{\rm seg} > N_{\rm thr}$.

This method ensures we remove very sparse data, whilst retaining $>90$ per cent of the data. In Figure~\ref{fig:lightcurves}, the first target in the $9<G<12$ group has 3 segments of data, where the middle segment contains relatively very few data points. We found that the gap between the first and second segment does not represent the downlink, but actually a time period where several data anomaly flags were returned.

\subsubsection{The detrending process}\label{sec:detrending}

As mentioned in $\S$\ref{sec:time_segments}, the (flux-normalised) TESS lightcurves often exhibit flux gradients across each segment (due to the instrument response), and clear discontinuities at the start/end points between neigbhouring segments. If the lightcurves are left like this, there is a strong risk that an LSP analysis (or any other $P_{\rm rot}$ measuring method) could calculate an incorrect, usually much longer $P_{\rm rot}$ value, because the jump in flux from one segment to the next may result in significant power output at lower frequencies during the Fourier transform. For these lightcurves, detrending is a crucial step.

\begin{figure}
    \centering
    \includegraphics[width=0.45\textwidth]{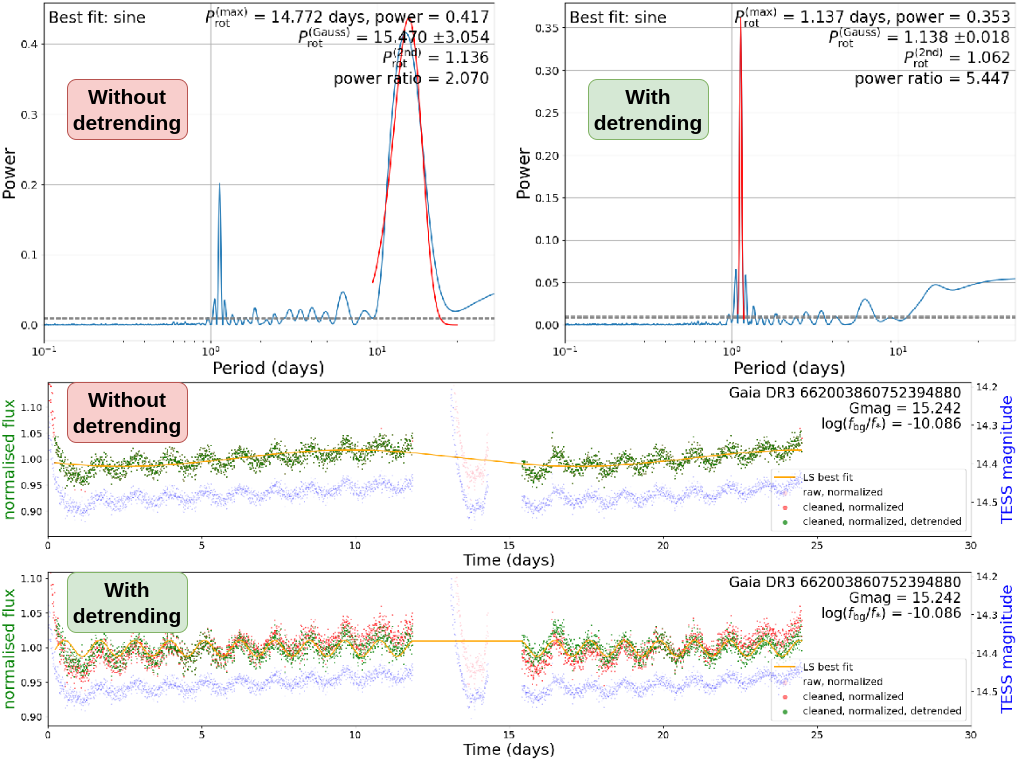}

    \vspace{1.0cm}
    \includegraphics[width=0.45\textwidth]{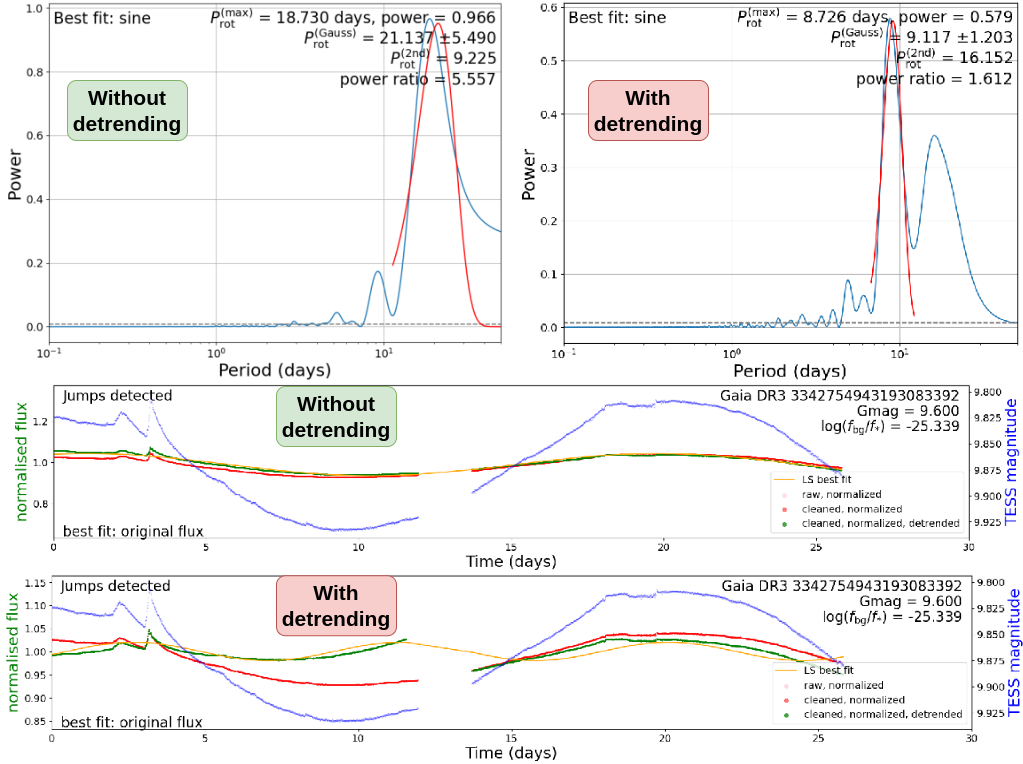}
    \caption{The effects of the detrending process on the LSP for two different lightcurves. The lightcurve in the top panel (TIC~302923006, Figure~\ref{fig:lightcurves}, $G>15$ sample, second panel) shows a $P_{\rm rot}=1.1\,$d is easy to confirm by visual inspection. However, because of the flux discontinuities between sectors, without any detrending, the LSP analysis predicts a longer $P_{\rm rot}$ ($\sim 15\,$d, left panel). When the detrending process is included, the expected $P_{\rm rot}$ value is recovered (right panel). The bottom panel represents a lightcurve for a target (HD~251108) where detrending would lead to an incorrect $P_{\rm rot}$. Note that this target has been previously measured with $P_{\rm rot}=21.18$\,d \citep{2006a_Watson}.}
    \label{fig:detrend_example}
\end{figure}

A clear example of when detrending is necessary is presented in the top panels of Figure~\ref{fig:detrend_example}~for TIC~302923006 (Gaia DR3 662003860752394880, see Figure~\ref{fig:lightcurves}, $G>15$ group, second panel). From visual inspection of the lightcurve, this target clearly has a $P_{\rm rot}$ value around $\sim 1.1\,$d, however, without the detrending step, the LSP analysis returns a much longer $P_{\rm rot}$ value ($\sim 14.8\,$d).

However, detrending does not always lead to improved lightcurves. In some (rare) cases, for targets with longer $P_{\rm rot}$ (e.g., $>10$\,d), the lightcurves may have sufficiently small systematic offsets during data gaps and the lightcurve is better without any detrending. An example of this is shown in the lower panels of Figure~\ref{fig:detrend_example}~for HD~251108, which displays a clear $P_{\rm rot}\sim20\,$d (this target has $P_{\rm rot}=21.18\,$d in the AAVSO International Variable Star Index VSX catalog, \citealt{2006a_Watson}). Because of the different ways that a lightcurve might be manifested, \TESSILATOR~runs several routines to determine if a lightcurve should be detrended.

For our detrending function, we make a polynomial fit of the form $f_{\rm fit} = \sum^{N}_{0} p_{n}t^{n}$, for $N=0$ and $N=1$. The value of $N$ is chosen using the Aikake Information Criterion (AIC) by comparing the sum of least squares for both fits. If the AIC value for $N=0$ is smaller, then this polynomial is used for detrending, otherwise the polynomial fit for $N=1$ is used. The reason for rejecting any higher order polynomials is because they would flatten the signal in lightcurves if the fit traced the shape the lightcurve, and also because they risk injecting periodocity in flat lightcurves (e.g., those that do not exhibit any periodic signal in the first place).

\vspace{0.5cm}
{\noindent\bf Step 1 -- smoothing}

The first part of the detrending process is to decide how to detrend the lightcurve (if at all), which begins with determining if the lightcurve has properties like HD~251108 (meaning no detrending should be performed at all). The idea is to make a sinusoidal fit to the data, and measure how good the fit is. The 5 steps for this process use parameters that have been selected through a trial and error process to maximise true positives and false negatives. They are:

\begin{itemize}
  \item [(1)] Detrend the whole lightcurve using a linear fit, and calculate the normal-scaled median absolute deviation (herein, simply referred to as ``MAD'') of the detrended flux values ($f_{\rm MAD}$).
  \item [(2)] Fit a sinusoidal function to the detrended lightcurve of the form $f_{\rm norm} = f_{0} + A\sin\left(\frac{2\pi}{P_{\rm rot}}t + \phi \right)$, where the parameters to be fit are $f_{0}$ (a constant), $A$ (the amplitude), $P_{\rm rot}$ (the rotation period) and $\phi$ (the phase) and $t$ represents the time coordinate.
  \item [(3)] Smooth the time and flux arrays using a boxcar of 10 data points (this is to remove short timescale irregular features such as flares).
  \item [(4)] Subtract the smoothed lightcurve by the sine fit and calculate the MAD ($d_{\rm MAD}$).
  \item [(5)] Calculate the relative root mean squared error (RRMSE) between the detrended flux and the sine fit ($f_{\rm RRMSE}$).
\end{itemize}

If the period from the sine fit is $>15\,$d, $f_{\rm RRMSE} < 0.01$ and $d_{\rm MAD}/f_{\rm MAD} < 0.25$, then the lightcurve is considered to be like HD~251108 and no detrending is made to the lightcurve. The output from this step is a boolean flag called ``smooth\_flag''.

\vspace{0.5cm}
{\noindent\bf Step 2 -- normalisation choice}
If ``smooth\_flag'' is False, then neighbouring segments of the lightcurve are assessed to determine if they are significantly offset, or if they would suitably connect. Another boolean flag called ``norm\_flag'' is initially set to False. Treating two consecutive segments $A$ and $B$, we provide a least-squares linear fit to both ($S_{A}$ and $S_{B}$), and calculate the predicted (extrapolated) flux value from $S_{A}$ at the first data point of $B$ ($B_{0}$), and compare this with the predicted value from $S_{B}$. We then detrend $A$ and $B$ (using $S_{A}$ and $S_{B}$) and calculate the MAD flux for each. If the absolute difference between $S_{A}$ and $S_{B}$ at $B_{0}$ ($|\Delta AB|$) is greater than twice the mean MAD value ($m$), then ``norm\_flag is set to True, and all segments of the lightcurve are detrended individually and the routine is stopped. Otherwise if there are $>2$ segments, the process is repeated for pairs $B$ and $C$, and so on. The lightcurve is detrended as a whole only if all pairs satisfy the condition that $|\Delta AB| < m$.

\subsubsection{Cleaning the segment edges: outliers}\label{sec:cleaning_outliers}
There are sometimes cases at the start and end parts of each segment where the flux values increase or decrease by an amount much larger than the typical scatter in the rest of the lightcurve. This is likely due to temperature changes in the detectors when they are switched off/on during data downlink. These often contribute a significant amount of noise, and removing these data points should lead to a cleaner lightcurve.

Therefore \TESSILATOR~uses a basic algorithm to remove the flux ``tails''. After the detrending processes are complete, the flux values for each segment are passed to the segment edge cleaning routine, and the median and MAD flux is calculated. Then from either edge of the segment (moving forwards and backwards in time from the start and end points, respectively), we check if the absolute difference in flux between the data point and the median is less than twice the MAD value. If not, the data point is rejected, and the test continues until a data point satisfies the condition. A clear example of this algorithm working in practice can be seen in the third panel of the $12<G<15$ group in Figure~\ref{fig:lightcurves}, where several data points at the end of each segment are removed (displayed as grey data points) because they have a sharp rise in flux.

\subsubsection{Cleaning the segment edges: scattered data}\label{sec:cleaning_scatter}
As well as the flux tails described in $\S$\ref{sec:cleaning_outliers}, the scatter amongst local flux values in regions close to the segment edges are sometimes significantly more scattered in relation to the rest of the segment. This may be a result of the detectors stabilising as the observations begin and/or end. Scattered data at the edges of segments can potentially contribute another noise term to the lightcurve, and removing them should lead to a cleaner lightcurve.

\TESSILATOR~provides a method to remove the scattered data by comparing the localised scatter with the rest of the segment. The method was developed with the intention to remove extreme data points, without removing potentially useful data. The idea is to select $N$ data points and calculate a running mean of the MAD values across the segment, where the value for $N$ is chosen as the lower value of either 0.01 times the number of data points in the segment or 11. If $N$ is even, we add one to it, so we can pad both edges of the running mean array with $(N-1)/2$ data points (the padded values are twice the MAD flux for the whole segment). We will denote this array as $f_{\rm mean}$.

The median absolute difference in flux between 2 neighbouring data points across the whole segment is calculated, which we denote as $f_{\rm diff}$. Then, from either edge of the segment (moving forwards and backwards in time from the start and end points, respectively) we test the condition $f_{\rm mean} < 2f_{\rm diff}$. If the condition is not satisfied, we repeat for the next value of $f_{\rm mean}$ in the array until we have a match, which marks where we set the start and end points of the segment. If the condition is immediately satisfied, we assume the segment edges have no significant extra scatter and no cleaning is performed. The same target used to demonstrate the edge outlier cleaning method ($\S$\ref{sec:cleaning_outliers}) can also be used to show the edge scatter cleaning method. At the end of the first segment it is clear that additional data points are removed because of large localised scatter. An example of the removal of highly scattered data at the edges of lightcurve segments can be seen in the second panel of the $12<G<15$ group in Figure~\ref{fig:lightcurves}.

\subsubsection{Removing extreme outliers}\label{sec:cleaning_extreme_outliers}
At this stage in the lightcurve analysis, the data has been normalised, segmented, detrended and cleaned. We then remove any remaining data points with flux values that can be considered as extreme outliers. To do so, we measure the median and MAD flux, and remove data points with flux values that are $>5$ MAD from the median flux. We note that the datapoints classed here as extreme outliers may be astrophysical in nature, for example optical flares \citep{2020a_Guenther} and/or accretion-driven bursts in young T-Tauri systems \citep{2023a_Lin}. In future \TESSILATOR~releases we plan to incorporate lightcurve smoothing functions, from which the residuals could be used to identify flares and classify flare properties.

\subsubsection{Final normalisation step}\label{sec:final_normalisation}
The final step is to divide the detrended flux in each segment by the median detrended flux for all data points in the segment that pass the cleaning steps in $\S$\ref{sec:cleaning_sparse_data}, $\S$\ref{sec:cleaning_outliers}, $\S$\ref{sec:cleaning_scatter} and $\S$\ref{sec:cleaning_extreme_outliers}. This concludes the processing steps for the lightcurve.

\subsubsection{Returning the processed lightcurve}\label{sec:lightcurve_finished}
The lightcurve processing method generates two products: a table of lightcurve data to be passed to further steps such as the LSP analysis (see $\S$\ref{sec:periodogram_analysis}) and a dictionary of parameters calculated in the detrending process ($\S$\ref{sec:detrending}). Both these products can be saved to file.

If the CBV-correction method is selected (see $\S$\ref{sec:CBV_corrections}) then the two products are made for both the original and the CBV-corrected fluxes. The lightcurve table retains information for all data points, regardless of whether or not they pass the 4 cleaning steps in the lightcurve analysis. The first 4 columns are the time stamp, TESS $T$ magnitude, input flux and the input flux error (directly taken from the results in $\S$\ref{sec:aperture}). Then the subsequent columns are the normalised flux and error (from $\S$\ref{sec:normalisation}), an index number used to group the segments ($\S$\ref{sec:time_segments}), a qualification flag for sparse data ($\S$\ref{sec:cleaning_sparse_data}), the detrended flux ($\S$\ref{sec:detrending}), and qualification flags from cleaning the segment edges from outliers ($\S$\ref{sec:cleaning_outliers}) and scatter ($\S$\ref{sec:cleaning_scatter}) and the extreme outlier removal ($\S$\ref{sec:cleaning_extreme_outliers}), and the median-divided final flux to be used for LSP analysis ($\S$\ref{sec:final_normalisation}).

\subsubsection{Additional lightcurve quality flags}\label{sec:flags}
Once the lightcurve has passed through the processing steps, 2 optional procedures are offered, both of which return a boolean flag that classify some additional lightcurves feature. The first tests if there appear to be significant discontinuities amongst neighbouring data points in a lightcurve segment. This might be a result of the instrumental response at the ends of each segment, but could also indicate a flare event or some other activity-related phenomenon. Therefore, we provide a ``jump\_flag'' for each lightcurve, which may prove useful for users depending on their reasons for using \TESSILATOR. Our flagging algorithm simply constructs an array of the running average of 10 consecutive flux values, which we label ``$A_{\rm run}$'', and another array containing the absolute difference in neighbouring $A_{\rm run}$ values, which we label ``$A_{\rm diff}$''. A ``jumpy'' lightcurve is one in which the ratio of the maximum and median values in $A_{\rm diff}$ becomes greater than 10 at any point across the segments.

The second procedure assesses whether the lightcurve is best matched to a sinusoidal or a linear fit. If the former is found, then the rotational signal likely strong enough to be detected, whereas the latter case suggests that the lightcurve is flat or noisy. Since there are 3 fixed parameters in the sinusoidal fit compared to 2 in the linear fit, the Aikaike Information Criterion is used to decide which of the least-squares fit is the most appropriate.

\subsubsection{Noise corrections using real data}\label{sec:noise_corrections}

We provide an alternative method for correcting instrumental systematics, by using the neighbouring targets identified as potential contaminants in $\S$\ref{sec:contamination}. For the method to run, the contamination calculation must be performed and there needs to be at least one contaminant identified in the image area. Each potential contaminant is passed through the same procedures as the target, and for each time stamp, a  median-flux value is calculated from all contaminants that satisfy all qualification criteria in $\S$\ref{sec:lightcurve_analysis}. The corrected target flux is calculated by dividing the final flux by the median contaminant flux at each time stamp.

This process is most effective when there are lots of nearby potential contaminants that have flat lightcurves, and in practice it is up to the user to decide if and when to apply this correction. 

\subsection{Periodogram Analysis}\label{sec:periodogram_analysis}
Whilst there are many methods to measure $P_{\rm rot}$ from photometric data, such as the autocorrelation method (ACF), wavelet transformations and Gaussian processes (to name just a few, see $\S$\ref{sec:introduction}), \TESSILATOR~only uses the Lomb-Scargle periodogram (LSP) method\footnote{We have experimented with using ACF, and we are considering providing ACF and/or other methods as an option in a future version of the software.}. The basic procedure for the LSP is to find a best least-squares sinusoidal fit to a lightcurve, using a grid of frequencies. Since LSP uses a simple linear regression to fit sinusoids, the computation time can be relatively fast, especially when Fast Fourier Transform algorithms are applied and an informed choice of frequency grid spacing and sampling factors is chosen. It is capable of measuring $P_{\rm rot}$ with unevenly sampled data (unlike ACF methods), and the periodogram output (a normalised power spectrum as a function of $P_{\rm rot}$) encodes important information across a wide range of input $P_{\rm rot}$ values, allowing us to investigate not just the $P_{\rm rot}$, but other sources of variability (e.g., accretion, magnetic activity, spot modulation).

The $P_{\rm rot}$ value corresponding to the maximum power is not necessarily the true $P_{\rm rot}$, for (at least) 2 reasons. The first involves the details of spot modulation. The angular resolution of TESS is far too large to resolve any surface features from a target. Therefore the cause of the flux variations is unknown, but generally assumed to be the migration of spots across the stellar disk. Although we assume a constant rotational speed for the spots, the monitoring of Sunspots \cite[e.g.][]{1863a_Carrington}, and spots on Solar-like stars \cite[e.g.][]{2017a_Brun} show a clear latitude dependency, with the fastest and slowest spots found at the equator and poles, respectively. This means the measured $P_{\rm rot}$ may be erroneous, or even meaningless, if a star has significant differential rotation. If the migration of spots from the Northern and Southern hemispheres are coherently out of phase as they move across the disk, the measured $P_{\rm rot}$ can be mistaken as half the true $P_{\rm rot}$, i.e., the spots are double-counted.

The second reason comes from the way the LSP deals with noisy and/or flat lightcurves. For these lightcurves the LSP may falsely yield a large power output at long $P_{\rm rot}$ (typically $> 10\,$d). This is mainly because small, gradual variations in flux across long timescales (e.g., across the entire sector) will have a higher sensitivity in the LSP, especially if there is no/little variability in the lightcurve on short timescales. In many of these cases, we can use some of the lightcurve properties and results from the LSP to reject them as poor data or at least flag them as spurious results. These ``quality indicators'' are described in $\S$\ref{sec:outputs_and_finalprot}.

\subsubsection{Implementing the Lomb-Scargle Periodogram calculation}\label{sec:LSP_implementation}
We use the {\sc LombScargle} python module
\citep{2022a_Astropy_Collaboration} 
with the auto-power method selected as an optimal comprise between accuracy and computational speed. An important step is to sample a frequency (or $P_{\rm rot}$) grid that is relevant for TESS lightcurves. With this in mind we select boundaries that correspond to $P_{\rm rot}=0.05$\,d and $100$\,d, respectively. The rationale for the lower limit is that the TESS/FFIC observing cadence is $\sim 0.007-0.021$\,d ($\sim 10-30$\,min, depending on the TESS mission mode during data acquisition). This ensures our frequency sampling rate is well below the Nyquist limit. In addition, low-mass stars are very rarely observed with $P_{\rm rot} < 0.2\,$d \citep{2013a_McQuillan,2020a_Ramsay}, therefore this lower limit ensures we can detect ultra-fast rotators. The upper limit was chosen because the duration of a TESS sector is $\sim 27$\,d, therefore reliable $P_{\rm rot}$ measurements are typically limited to $<10-12$\,d. However, we set our upper limit at $P_{\rm rot}=100\,$d because we need to fit error bars for lightcurves with longer $P_{\rm rot}$. Setting a sampling rate of 10 points for each frequency node results in a grid of $\sim 6\,000$ points. The typical spacing in $P_{\rm rot}/d$ is $\sim 0.003, 0.1$ and $0.4$ at $P_{\rm rot}/d$ values of $1, 5$ and $10$, respectively.

The data selected for the LSP are described in $\S$\ref{sec:lightcurve_finished}, with the time, flux and flux error provided as input, which will herein be denoted simply as $t$, $f$ and $\sigma_{f}$, respectively. After running the LSP, the $P_{\rm rot}$ and normalised power output (herein simply ``power'') are recorded, and saved to file if necessary. Using the value of the maximum power output, False Alarm Probability (FAP) powers are then calculated assuming a null hypothesis of non-varying data with Gaussian noise at the 0.1, 1.0 and 5.0 per cent level, based on the method described by \cite{2008a_Baluev}.

\subsubsection{Fitting Gaussian profiles to the highest {\it n} peaks}\label{sec:highest_n_peaks}
In $\S$\ref{sec:periodogram_analysis}~we explained why the highest power output of the LSP may not represent the true $P_{\rm rot}$. Therefore we designed a method to return $P_{\rm rot}$ values from the highest $n$ power outputs from distinct peaks in the LSP, where we set $n=4$.

The method works by identifying the array index corresponding to the maximum power output, and counting all neighbouring preceding and subsequent indices with power values that decrease monotonically until there is an inflection and the power value increases. The LSP power spectra for lightcurves that are noisy or possessing multiple components can sometimes have peaks that are asymmetric, and the top of these peaks can be double-humped. Therefore we make an exceptional rule that if a data point has a power value $> 85\,$per cent of the maximum power, we allow the program to continue running even if the new power value represents an increase.

The $P_{\rm rot}$ and power values between the start and the end index from this method are used to make a Gaussian fit. The centroid and full-width half maximum are used to provide a Gaussian $P_{\rm rot}$ and uncertainty. This information is saved, and these parts of the LSP are removed. The process is then repeated $n$ times for the desired number of LSP peaks.

\subsubsection{The shuffled period method}\label{sec:shuffled_period}
The general LSP method, as described in $\S$\ref{sec:LSP_implementation}, uses lightcurve data measured over a full TESS sector. As we have discussed, a reliable $P_{\rm rot}$ measurement necessitates a lightcurve in which the stellar rotation provides the dominant signal. Despite considerable normalisation, detrending and cleaning steps (see $\S$\ref{sec:lightcurve_analysis}), signals from the systematics, sky background and non-periodic surface features may dominate those from the underlying rotation, especially for faint sources with low signal to noise.

A visual inspection of some \TESSILATOR-produced lightcurves for faint sources with short $P_{\rm rot}$ ($\lesssim 3\,$d, as measured by {\it Kepler-K2}, \citealt{2014a_Howell}) show that the rotational signal is undulating and apparent. However the LSP for these targets often return relatively little power at the correct $P_{\rm rot}$, because other components -- usually with much lower frequency -- dominate the signal.

We have developed a specialised LSP method, specifically designed to measure $P_{\rm rot}$ in cases where the amplitude of the rotational signal is too small to be detected when the full sector of lightcurve data is used. The general idea is to randomly sample many lightcurve subsets across the sector and measure their $P_{\rm rot}$. If the distribution of $P_{\rm rot}$ values from these randomly sampled subsets has a small-enough variance, then its mean value, which we denote as $P_{\rm shuff}$, replaces the $P_{\rm rot}$ corresponding to the LSP peak power output across the full sector. Specifically, there are 5 steps to the method:

\begin{enumerate}
    \item [(1)]{\bf Measuring $P_{\rm rot}$ for many subsets of the lightcurve.}\\
    We select 5000 randomised subsets of lightcurve data from the whole sector. The data for each subset is chosen by: (a) randomly selecting a lightcurve segment (see $\S$\ref{sec:time_segments}); (b) extracting a random fraction, $f$, of the chosen segment, between 0.1 and 1; and (c) selecting the starting point of the subset as a random fraction between 0 and $1-f$ of the segment. Each subset is linearly de-trended, and the LSP is performed as described in $\S$\ref{sec:LSP_implementation}, but with a maximum $P_{\rm rot}$ set to 4 times the duration of the subset. This yields 5000 sampled $P_{\rm rot}$ measurements (herein $P_{\rm S}$).

    \item [(2)]{\bf Constructing a histogram of $P_{\rm S}$.}\\
    The first histogram ($H_{1}$) represents a broad fit of all $P_{\rm S}$ values, which is used to find the $P_{\rm rot}$ bounds of the distribution surrounding the most populated bin. The $P_{\rm S}$ are placed onto a base-10 logarithmic scale, and split into 50 equal-spaced bins.

    \item [(3)]{\bf Identifying the $P_{\rm rot}$ range surrounding the peak of $H_{1}$.}\\
    Using the algorithm described in the second paragraph of $\S$\ref{sec:highest_n_peaks}~(here we replace the power output with the number occupancy in each bin and we set the ``exception rule'' as 5 times the median number occpancy), we measure $P_{S_{L}}$ and $P_{S{_U}}$, which are the $P_{S}$ lower and upper values that represent the bounds of the region surrounding the maximum occupancy value. The fraction of the total $P_{\rm S}$ that lie within $P_{S_{L}}$ and $P_{S{_U}}$ (which we label $f_{1}$) is calculated, and a second, refined histogram ($H_{2}$) is produced, this time using 10 times the number of bins that (at least partially) lie within $P_{S_{L}}$ and $P_{S{_U}}$, normalised such that they summate to 1.

    \item [(4)]{\bf Applying the first set of criteria.}\\
    The criteria that must be matched for $P_{\rm shuff}$ to remain in consideration are: (a) the number of $H_{2}$ bins that lie entirely within $P_{S_{L}}$ and $P_{S{_U}}$ must be $> 3$; (b) $f_{1} > 0.5$; and (c) the $H_{2}$ peak must not be in a bin that overlaps with $P_{S_{L}}$ and $P_{S{_U}}$. The process only continues if these conditions pass, otherwise the $P_{\rm shuff}$ is rejected.

    \item [(5)]{\bf Applying the second set of criteria.}\\
    We make a Gaussian fit to the binned data in $H_{2}$, and measure the relative root mean square error (RRMSE). At this stage we calculate the value of $P_{\rm shuff}$ and uncertainty from the Gaussian fit parameters. The second set of criteria that must be matched in order to replace $P_{\rm rot}$ with $P_{\rm shuff}$ are: (a) RRMSE $<0.5$; (b) the full-width half maximum of the Gaussian fit $<0.05$; and (c) $|1-(P_{\rm shuff}/P_{\rm rot})|>0.1$.
\end{enumerate}

\begin{figure}
    \centering
    \includegraphics[width=0.48\textwidth]{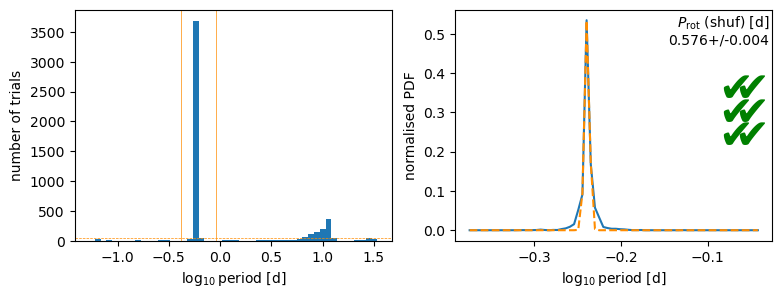}
    \includegraphics[width=0.48\textwidth]{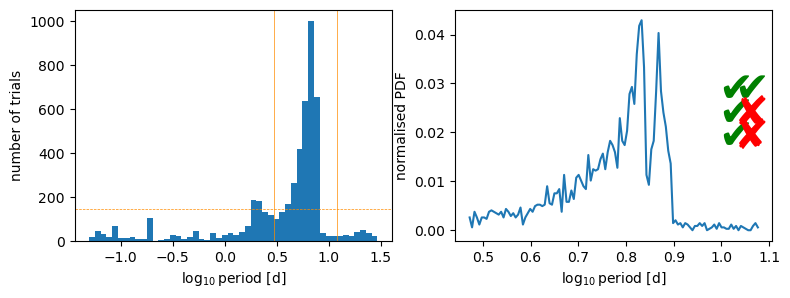}
    \caption{An example of the results of the $P_{\rm rot}$ histograms for a case where the shuffled $P_{\rm rot}$ is selected (top panel) and another where some criteria fail, and therefore $P_{\rm shuff}$ is not used (see $\S\ref{sec:shuffled_period}$ for details). The left panels show the initial $P_{\rm S}$ distribution, where the orange horizontal line indicates $5\times$ the median bin occupancy, and the orange vertical lines show the resulting $P_{S_{L}}$ and $P_{S_{U}}$ bounds (see step 3 in the routine) that are used to construct the refined $P_{\rm shuff}$ histogram, shown on the right. The target in the upper panel satisfies all 6 criteria and the target in the lower panel fails 2 criteria (b and c in step 5).}
    \label{fig:shuff_period}
\end{figure}

Figure~\ref{fig:shuff_period}~provides an example of when $P_{\rm shuff}$ is, and is not adopted. Generally, if there is a clear, rapidly rotating $P_{\rm shuff}$, then the scatter in $P_{S}$ is very small, meaning the number of false positives is very low. We note that this method is particularly effective at detecting $P_{\rm rot}$ in lightcurves with low-amplitude for rapid rotators, or at least $P_{\rm rot}$ shorter than the duration of a typical segment. However, the shuffle method is not capable of finding longer $P_{\rm rot}$, given that the $P_{\rm shuff}$ values are never longer than the time duration of a segment.

\subsubsection{Phase-folded lightcurves}\label{sec:phase_folding}

A phase-folded lightcurve (PFL) is formed by simply dividing the time coordinates by the selected $P_{\rm rot}$. Displaying the lightcurve in this format allows one to visualise repeating patterns in a $P_{\rm rot}$ cycle such as rotational modulation, transits, eclipses or accretion events, which may be less easy to identify in the full lightcurve. Each time coordinate is zero-shifted (such that the first datapoint is equal to zero), and then divided by $P_{\rm rot}$. The whole part and decimal part of these transformed values represent the number of cycles and the phase, respectively.

We measure four quantities from the PFL, which are used to help assess the reliability of the $P_{\rm rot}$ measurement. The first two of these are a measure of the amplitude ($A$) and scatter ($s$) of a sinusoidal fit to the PFL, of the form $f_{\rm fit}=A\sin(2\pi t + \phi) + f_{0}$, where $\phi$ and $f_{0}$ represent the phase and a constant, respectively. To measure $s$, we calculate a value $\rho=f/f_{\rm fit}$, then split the PFL into ten equal phase sections and calculate the mean ($\overline{\rho}$) and standard deviation ($\sigma_{\rho}$) for each section. Finally, $s$ is given as the median $\sigma_{\rho}$ over the ten sections. The third quantity is a reduced $\chi^{2}$ score for the sinusoidal fit, and the fourth is the fraction of data points with $|f-f_{\rm fit}|>3 \times f_{\rm MAD}$, which we deem as extreme outliers.


\subsection{Outputs, quality indicators and calculating a final $P_{\rm rot}$}\label{sec:outputs_and_finalprot}
\begin{figure*}
    \centering
    \includegraphics[width=0.9\textwidth]{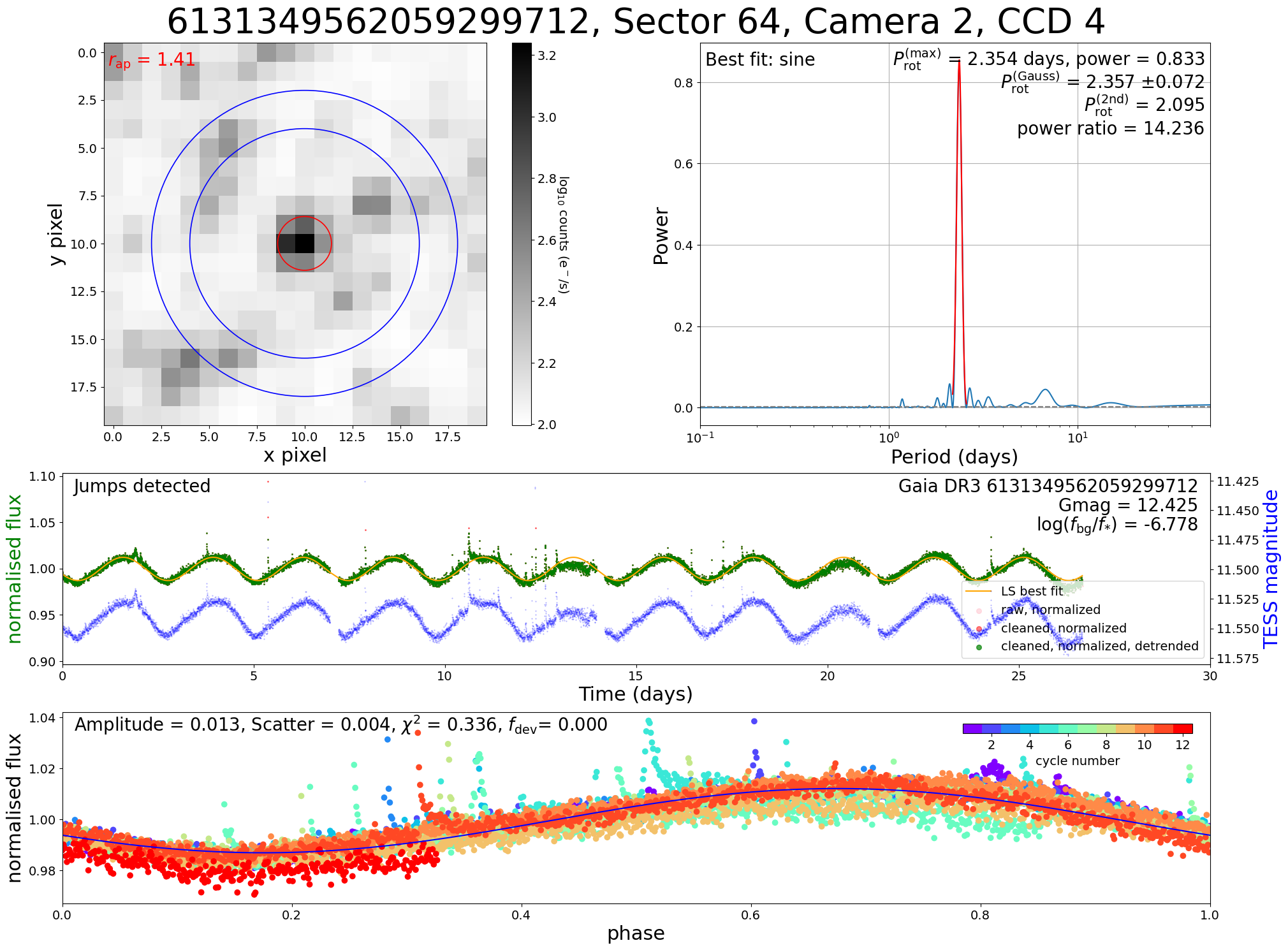}
    \caption{A plot generated with \TESSILATOR, which displays the main results of the processing steps described in $\S$\ref{sec:tessilator_code}. The title gives the original\_id and SCC. Top-left: a $20\times20$ pixel cutout from the middle slice of the TESS image stack, with the greyscale representing the range of base-10 logarithmic counts. The red inner circle represents the aperture radius (the value is given in the top left) and the blue circles are the annuli used to calculate the background. Middle: the lightcurve created by \TESSILATOR, where blue points represent the TESS magnitudes calculated during aperture photometry and pink, red and green points represent the normalised fluxes in: their original form; after detrending; and those that pass the cleaning processes. The orange line represents a best-fit sinusoid, calculated during the LSP analysis. Top-right: Results from the LSP analysis, where the red section indicates the region around the selected $P_{\rm rot}$. The dashed horizontal line near Power=0.0 represents the power output of the FAP at 1 per cent. Bottom: the green lightcurve from the middle panel phase-folded using the selected $P_{\rm rot}$. A rainbow colormap indicates the cycle of each phase throughout the sector, and the blue curve represents the best-fit sine function to the phase-folded lightcurve.}
\label{fig:tessilator_summary_plot}
\end{figure*}

Once \TESSILATOR~has completed the procedures in $\S$\ref{sec:target_input_data_retrieval}, $\S$\ref{sec:aperture}, $\S$\ref{sec:lightcurve_analysis}~and $\S$\ref{sec:periodogram_analysis}~(see Figure~\ref{fig:flowchart}), a results table for all measured parameters of a given lightcurve is returned, which may be saved to file. Table~\ref{tab:tessilator_outputs}~provides details for each of the columns of this results file, which consist of general Gaia and TESS demographic information, details on the background contamination, boolean flags from the lightcurve vetting, and outputs from the periodogram analysis and phase-folding. If required, a plot summarising the main results is provided. An example of a \TESSILATOR~summary plot is shown in Figure~\ref{fig:tessilator_summary_plot}.

\begin{table}
\begin{tabular}{ll}
\hline\hline
Column name & Description \\
\hline
\multicolumn{2}{c}{General properties} \\

\hline
original\_id & Target identifier ($\S\ref{sec:target_input}$)  \\
source\_id & Gaia DR3 source identifier \\
ra & Right ascension (epoch J2000) \\
dec & Declination (epoch J2000) \\
parallax & Gaia DR3 parallax \\
Gmag & Gaia DR3 $G$-band magnitude \\
BPmag & Gaia DR3 $G_{\rm BP}$-band magnitude \\
RPmag & Gaia DR3 $G_{\rm RP}$-band magnitude \\
Tmag\_MED & Median TESS $T$-band magnitude ($\S$\ref{sec:aperture}) \\
Tmag\_MAD & MAD TESS $T$-band magnitude ($\S$\ref{sec:aperture}) \\
Sector & TESS sector number \\
Camera & TESS camera number \\
CCD & TESS CCD number \\

\hline
\multicolumn{2}{c}{Background contamination} \\
\hline
log\_tot\_bg & $\Sigma\eta$ ($\S\ref{sec:contamination}$) \\
log\_max\_bg & $\eta_{\rm max}$ ($\S\ref{sec:contamination}$) \\
num\_tot\_bg & Number of contaminating sources \\
false\_flag & Test if a contaminant is the $P_{\rm rot}$ source ($\S\ref{sec:contamination}$) \\
reliable\_flag & Test if the $P_{\rm rot}$ source is reliable ($\S\ref{sec:contamination}$) \\

\hline
\multicolumn{2}{c}{Lightcurve properties} \\
\hline
CBV\_flag & The CBV-correction category ($\S\ref{sec:CBV_corrections}$) \\
smooth\_flag & Flag for detrending step 1 ($\S\ref{sec:detrending})$ \\
norm\_flag & Flag for detrending step 2 ($\S\ref{sec:detrending})$ \\
jump\_flag & Test for jumps in the lightcurve ($\S\ref{sec:flags}$) \\
AIC\_line & AIC score: linear fit to the lightcurve ($\S\ref{sec:flags}$) \\
AIC\_sine & AIC score: sine fit to the lightcurve ($\S\ref{sec:flags}$) \\

\hline
\multicolumn{2}{c}{Periodogram analysis} \\
\hline
Ndata & Number of datapoints in the periodogram analysis \\
FAP\_001 & 1\% False Alarm Probability power ($\S\ref{sec:periodogram_analysis}$) \\
period\_1 & Primary $P_{\rm rot}$ (peak) \\
period\_1\_fit & Primary $P_{\rm rot}$ (Gaussian fit centroid) \\
period\_1\_err & Primary $P_{\rm rot}$ uncertainty \\
power\_1 & Power output of the primary $P_{\rm rot}$ \\
period\_2 & Secondary $P_{\rm rot}$ (peak) \\
period\_2\_fit & Secondary $P_{\rm rot}$ (Gaussian fit centroid) \\
period\_2\_err & Secondary $P_{\rm rot}$ uncertainty \\
power\_2 & Power output of the secondary $P_{\rm rot}$ \\
period\_3 & Tertiary $P_{\rm rot}$ (peak) \\
period\_3\_fit & Tertiary $P_{\rm rot}$ (Gaussian fit centroid) \\
period\_3\_err & Tertiary $P_{\rm rot}$ uncertainty \\
power\_3 & Power output of the tertiary $P_{\rm rot}$ \\
period\_4 & Quaternary $P_{\rm rot}$ (peak) \\
period\_4\_fit & Quaternary $P_{\rm rot}$ (Gaussian fit centroid) \\
period\_4\_err & Quaternary $P_{\rm rot}$ uncertainty \\
power\_4 & Power output of the quaternary $P_{\rm rot}$ \\
period\_shuffle & $P_{\rm shuff}$ ($\S\ref{sec:shuffled_period}$) \\
period\_shuffle\_err & Uncertainty in $P_{\rm shuff}$ ($\S\ref{sec:shuffled_period}$) \\
shuffle\_flag & Indicates if $P_{\rm shuff}$ was adopted ($\S\ref{sec:shuffled_period}$) \\

\hline
\multicolumn{2}{c}{Phase-folded lightcurve details} \\
\hline
amp & Amplitude of the PFL ($\S\ref{sec:phase_folding}$) \\
scatter & Scatter of the PFL ($\S\ref{sec:phase_folding}$) \\
chisq\_phase & $\chi^{2}$ value of the sinusoidal fit to the PFL ($\S\ref{sec:phase_folding}$) \\
fdev & Number of extreme outliers in the PFL ($\S\ref{sec:phase_folding}$) \\
\hline
\end{tabular}
\caption{The names and description of each column in the results table. When necessary, we refer the reader to the appropriate section in this manuscript for more details.}
\label{tab:tessilator_outputs}

\end{table}

\subsubsection{Quality indicators}\label{sec:tessilator_outputs}

The quality and reliability of a TESS lightcurve depends on many factors. These are typically based on: the flux ratio between the target and surrounding background and/or neighbouring contaminants; the instrumental response in a given SCC; and the physical parameters and surface features of the target. For these reasons, \TESSILATOR~measures several quantities that can be used to assess the quality and reliability of the lightcurve, as described in Table~\ref{tab:tessilator_outputs}. Here we mostly focus on the criteria used in $\S$\ref{sec:R23_comparison}~to vet the 6551 lightcurves used for the comparison with R23.

For the background contamination, we suggest flagging lightcurves with ``log\_tot\_bg\_star''$>0$ as potentially unreliable. This is because the background contaminants would represent at least half of the flux in the aperture, making it difficult to determine the source of any rotational modulation. Of course, it is possible to further study the effects of each contaminant source if required, using the ``false'' and ``reliable'' flags.

Although the lightcurve analysis provides quality criteria on whether the lightcurve contains jumps (``jump\_flag'') and the flatness (``AIC\_line'' and ``AIC\_sine''), we do not consider these as strong quality criteria. This is particularly important for the latter criteria, since it removes many faint targets with low SNR.

The outputs from the LSP provide several important quality indicators. The ratio between the power of the LSP peak and the power corresponding to the FAP at 1 per cent should be sufficiently large, the $P_{\rm rot}$ uncertainty should be relatively small, and long $P_{\rm rot}$ values should be treated with caution.

The details from the phase-folded lightcurve provide important criteria. The ratio of the amplitude and scatter can be set to a minimum threshold to avoid noisy, flat lightcurves. Additionally, if the amplitude or the scatter is very large, this may be indicative of a poor lightcurve. Finally, the ``chisq\_phase'' and ``fdev'' values can be set, such that they should not exceed maximum values.

The choice of quality criteria are largely subjective and depend on the requirements of the user. For example, if faint targets are being analysed, then the amplitude is likely to be low (so a criteria based on amplitude should be relaxed), or if a target is in a crowded field, one may wish to relax the criteria for contamination. In future work we plan to experiment further with these quality criteria using machine learning tools such as random forest classifiers to determine the relative importance of these features.

\subsubsection{Calculating a final $P_{\rm rot}$}\label{sec:final_period}

The task of calculating a final $P_{\rm rot}$ (herein, $P_{\rm fin}$) for a target when there are lightcurves from 2 or more sectors is not trivial. Before we describe our method, we note that \TESSILATOR~does not measure differential rotation (nor indicate them). For small datasets we recommend that users visually inspect lightcurves and apply intuition to ensure reliable $P_{\rm fin}$ values. However, for large target samples, or even when there are many sectors of data for just one target, an automated method for measuring $P_{\rm fin}$ is preferable.

Our method is largely adopted from sections 3.1 and 3.2 in R23, which consists of two steps: determining the $P_{\rm rot}$ value to be selected from each individual lightcurve ($P_{\rm sec}$)  and calculating $P_{\rm fin}$ using all the $P_{\rm sec}$ values.

\vspace{0.5cm}
{\noindent\bf Step 1 -- Determining $P_{\rm rot}$ for a single lightcurve ($P_{\rm sec}$)}

For lightcurves with clear, unambiguous rotational modulation, the selected $P_{\rm sec}$ is usually the $P_{\rm rot}$ value corresponding to the highest peak of the LSP (herein $P_{\rm rot,1}$). However, this is not always the case. There two cases where an alternative $P_{\rm rot}$ may be selected for $P_{\rm sec}$. As described in $\S$\ref{sec:highest_n_peaks}, \TESSILATOR~provides $P_{\rm rot}$ for the highest $n$ peaks from the LSP, where the default is $n=4$.

The first case relates to slow rotators. For the purpose of our automated $P_{\rm sec}$ measurement, we remove any $P_{\rm rot}$ values $>20\,$d. Whilst this means our pipeline is unable to identify targets with $P_{\rm sec}>20\,$d, we find that TESS, with sectors only lasting $\sim 27\,$d, is usually incapable of measuring rotation slower than this. We find it more favourable in general to eliminate false positives in this way. In almost all cases, the TESS lightcurves for targets whose $P_{\rm rot}$ values are {\it known} to be $>20\,$d (from surveys with longer observing baselines) appear very flat, and therefore are eliminated due to some quality criteria failure (see $\S$\ref{sec:tessilator_outputs}).

The second case accounts for potential double-counting of spots, which can lead to estimates of $P_{\rm rot}$ that are half the true value \citep{2018a_Basri}. As in R23, we select an alternative $P_{\rm rot}$ from the $n\geq2$ set if this matches twice the $P_{\rm rot,1}$ value (within one error bar of the alternative $P_{\rm rot}$) and the power corresponding to this peak is $>0.6$ that of $P_{\rm rot,1}$.

\vspace{0.5cm}
{\noindent\bf Step 2 -- Determining the final $P_{\rm rot}$ value ($P_{\rm fin}$)}

The $P_{\rm sec}$ values, errors and corresponding power output values, selected in step 1, are used to calculate $P_{\rm fin}$. How $P_{\rm fin}$ is calculated depends on whether the number of $P_{\rm sec}$ values ($N_{\rm sec}$), is 1, 2, or $>2$.

For $N_{\rm sec}=1$, $P_{\rm fin}$ is obviously equal to $P_{\rm sec}$ and no further analysis is needed. For $N_{\rm sec}=2$, we first check if one $P_{\rm sec}$ is approximately double that of the other. If the difference between the longest $P_{\rm sec}$ and twice the shortest $P_{\rm sec}$ is less than either the mean of both error bars or 10 per cent of the shortest $P_{\rm rot}$, then $P_{\rm fin}$ is selected as the longest $P_{\rm sec}$. Otherwise, $P_{\rm fin}$ is the power-weighted mean of both measurements.

Determining $P_{\rm fin}$ for $N_{\rm sec} > 2$ is slightly more complex. Like in the $N_{\rm sec} = 2$ case, we need to check if any of the $P_{\rm sec}$ measurements in the set are double values. The first step is to measure the median $P_{\rm sec}$ ($P_{\rm med}$) and the corresponding error bar ($\sigma_{P_{\rm med}}$). Then we determine which lightcurves have either a half or full $P_{\rm sec}$ within one $\sigma_{P_{\rm med}}$ value of $P_{\rm med}$. The final chosen $P_{\rm fin}$ depends on whether more than two-thirds of these match this criterion.
\begin{itemize}
    \item If $\geq 2/3$ satisfy the criterion, only use sectors that represent matches. If there are no ``half'' matches in this set, then $P_{\rm fin}$ is the power-weighted mean. Otherwise, ensure the $P_{\rm sec}$ values match the double value (by doubling their $P_{\rm sec}$ and errors if necessary) and take $P_{\rm fin}$ as the median value.
    \item Otherwise, take $P_{\rm fin}$ as the power-weighted mean $P_{\rm sec}$ from all the individual $P_{\rm sec}$ values.
\end{itemize}

Selecting $P_{\rm fin}$ in this way allows us to utilise as much TESS data for multiple sectors, and also makes the code more likely to select the correct ``double value'' for $P_{\rm fin}$ in cases where there are measurements that are different by factors of 2. We reiterate that our goal here is to develop an automated method to calculate $P_{\rm fin}$ for large numbers of lightcurves. It works in most cases, but there will inevitably be some occasions where this method is not optimal. We advise the user to visually inspect lightcurves for best results.

\section{Comparison with literature samples}\label{sec:period_comparison}

\subsection{The Rampalli et al. 2023 sample}\label{sec:R23_comparison}

\subsubsection{Running \TESSILATOR~on the R23 sample}\label{sec:R23_tessilator_run}

To validate our software as a tool capable of correctly measuring periods, we run \TESSILATOR~on the sample of 1560 stars targeted by R23. A number of these R23 targets are used as exemplars to demonstrate the \TESSILATOR~procedures throughout $\S$\ref{sec:tessilator_code}. The R23 target list comprises members of 5 different groups: the Pisces-Eridani kinematic stream  (Pis-Eri); the Pleiades, Hyades and Praesepe open clusters; and the M-Earth sample of nearby field M-stars. The first 4 of these groups are $\sim 100-700$\,Myr, which is an epoch where we expect the majority of stars to rotate faster than $\sim10\,$d \citep{2003a_Barnes,2013a_Gallet,2019a_Angus} and therefore capable of having $P_{\rm rot}$ correctly measured with TESS data. The M-Earth sample provides a useful case study for targets that are likely to have a true $P_{\rm rot}>10\,$d. TESS lightcurves are available for at least one target across every TESS sector (from 1-72) and most targets have multiple sectors of data. The magnitude range of the full sample is $6 < G < 18$, and the spectral-type range spans early-A to late-M.

This makes the R23 sample ideal for testing \TESSILATOR's performance across a large parameter space. All targets in the R23 sample themselves have a $P_{\rm rot}$ comparison with a literature source (this set we denote as ``R23\_lit''), which we will additionally test the \TESSILATOR~with. In particular, the provenance of the literature $P_{\rm rot}$ for the 3 open clusters is the {\it Kepler-K2} survey, allowing for a comparative study between different satellite photometric missions. Details of each group are summarised in table 1 of R23.

To expedite our code, we ran \TESSILATOR~as a parallel-processing task, using the ``bwForCluster BinAC'' facility at the Baden-W\"urrtemburg Center for High Performance Computing\footnote{\url{https://uni-tuebingen.de/en/einrichtungen/zentrum-fuer-datenverarbeitung/dienstleistungen/server/computing/resources/bwforcluster-binac/}}. We set a limit of no more than 10 sectors of data per target and decided not to apply any CBV corrections. The run split the target list of 1560 equally among 90 cores, each of which was allocated 1Gb of RAM. The job completed in approximately 5 hours. The run produced output for 6551 lightcurves for 1543 targets (for 17 targets there are no lightcurves because \TESSILATOR~ran into either an SQL server issue, or the lightcurve analysis failed to run).

\subsubsection{Fixed \TESSILATOR~criteria for R23}\label{sec:R23_fixed_criteria}
Our first test is to determine if we should apply a set of fixed quality criteria to vet the lightcurves (see Table~\ref{tab:tessilator_outputs}~for a description of the parameters). If $P_{\rm shuff}$ is used (i.e., ``shuffle\_flag''=1), then no additional criteria are applied (except $P_{\rm shuff}>0.1$), otherwise we set the following criteria: 
\begin{itemize}
    \item log\_tot\_bg $<0$
    \item period\_1\_err/period\_1 $<0.25$
    \item $0.1<$period\_1$<20.0$
    \item chisq\_phase $<0.4$
    \item scatter $<0.1$
    \item amp $<0.2$
    \item fdev $<0.01$
\end{itemize}

With these criteria applied, we lose about 30 per cent of the lightcurves (1967 in total) and 116 targets, leaving us with 4557 lightcurves (687 with shuffle\_flag=1) and 1429 targets. 

Our aim is to find the number of targets that have matching $P_{\rm fin}$ values between \TESSILATOR~and R23/R23\_lit measurements. To do so, we measure the modulus of the difference between the period measurements, and assume that a successful match is achieved when this is less than either twice the uncertainty in $P_{\rm fin}$ or 0.1\,d, where the latter ensures a match for very fast rotators when the $P_{\rm fin}$ error is very small. The total fraction of matches across all 5 groups (which we label $f_{\rm tot}$), if the above criteria are, or are not implemented, is almost identical for the R23 and R23\_lit samples: $\sim 67$ per cent ($966/1429$ and $1034/1543$) and $\sim 62$ per cent ($912/1436$ and $949/1543$), respectively. Both results suggest the effect of applying the above criteria has negligible effect whatsoever on the lightcurve quality, but it reduces the number of stars that we get $P_{\rm rot}$ for.

\subsubsection{Variable \TESSILATOR~criteria for R23}\label{sec:R23_variable_criteria}

In general, enforcing more restrictive criteria should result in fewer targets, but a higher fraction of matches. To test this, we employ two further criteria, but this time we make the pass threshold variable. These are the ratios of ``amp'' and ``scatter'' (herein AS), and ``power\_1'' and ``FAP\_001'' (PF). Table~\ref{tab:match_criteria}~represents a grid of results for various critical values of each ratio. It generally appears that a higher AS and PF value improve $f_{\rm tot}$ for both samples by approximately 10 per cent, however, this comes at the cost of losing almost two-thirds of the initial sample. We see that that highest values of $f_{\rm tot}$ are actually somewhere in the range of $2.0<$AS$<3.0$, whereas PF has an almost negligible effect on $f_{\rm tot}$ except at very high values (i.e., $\geq 50$). A comparison of the R23 and R23\_lit results indicate that when PF$<50$, for AS$\leq 1$, $1<$AS$<3$ and AS$>3$, the R23\_lit $f_{\rm tot}$ are typically $\sim 0.01-0.04$ lower, $\sim 0.01-0.02$ higher, and about equal, respectively. These results suggest that AS and PF could be strongly correlated, therefore we drop PF as a criterion.

\begin{table*}
\begin{tabular}{lrrrrrr}
\hline\hline
AS & FP=0 & FP=1 & FP=5 & FP=10 & FP=50 & FP=100 \\
\hline
\multicolumn{7}{c}{Comparison with R23} \\
0.0 & 966/1429=0.68 & 960/1419=0.68 & 935/1363=0.69 & 912/1288=0.71 & 763/1028=0.74 & 573/771=0.74 \\
0.5 & 896/1267=0.71 & 896/1267=0.71 & 896/1265=0.71 & 891/1246=0.72 & 762/1028=0.74 & 573/771=0.74 \\
1.0 & 829/1095=0.76 & 829/1095=0.76 & 829/1095=0.76 & 829/1094=0.76 &  739/971=0.76 & 570/768=0.76 \\
1.5 &  766/990=0.77 &  766/990=0.77 &  766/990=0.77 &  766/990=0.77 &  720/923=0.78 & 526/679=0.77 \\
2.0 &  718/887=0.81 &  718/887=0.81 &  718/887=0.81 &  718/887=0.81 &  674/840=0.80 & 477/603=0.79 \\
2.5 &  654/797=0.82 &  654/797=0.82 &  654/797=0.82 &  654/797=0.82 &  619/761=0.81 & 448/566=0.79 \\
3.0 &  599/732=0.82 &  599/732=0.82 &  599/732=0.82 &  599/732=0.82 &  569/701=0.81 & 433/545=0.81 \\
3.5 &  550/674=0.82 &  550/674=0.82 &  550/674=0.82 &  550/674=0.82 &  525/648=0.81 & 420/530=0.79 \\
4.0 &  508/628=0.81 &  508/628=0.81 &  508/628=0.81 &  508/628=0.81 &  489/608=0.80 & 406/514=0.79 \\
\hline

\multicolumn{7}{c}{Comparison with literature sources in R23} \\
0.0 &  912 (-0.04) &   911 (-0.04) &   889 (-0.03) &   876 (-0.03) &    756 (+0.00) &  565 (-0.01) \\
0.5 &  877 (-0.02) &   877 (-0.02) &   877 (-0.02) &   874 (-0.02) &    756 (+0.00) &  565 (-0.01) \\
1.0 &  821 (-0.02) &   821 (-0.02) &   821 (-0.02) &   821 (-0.01) &    740 (+0.00) &  563 (-0.03) \\
1.5 &  784 (+0.02) &   784 (+0.02) &   784 (+0.02) &   784 (+0.02) &    738 (+0.02) &  525 (+0.00) \\
2.0 &  732 (+0.02) &   732 (+0.02) &   732 (+0.02) &   732 (+0.02) &    692 (+0.02) &  472 (-0.01) \\
2.5 &  661 (+0.01) &   661 (+0.01) &   661 (+0.01) &   661 (+0.01) &    627 (+0.01) &  441 (-0.01) \\
3.0 &  602 (+0.00) &   602 (+0.00) &   602 (+0.00) &   602 (+0.00) &    573 (+0.01) &  423 (-0.03) \\
3.5 &  550 (+0.00) &   550 (+0.00) &   550 (+0.00) &   550 (+0.00) &    525 (+0.00) &  409 (-0.02) \\
4.0 &  505 (-0.01) &   505 (-0.01) &   505 (-0.01) &   505 (-0.01) &    486 (+0.00) &  393 (-0.03) \\
\hline
\end{tabular}
\caption{Results for the period matching criterion when the variable criteria ``AS'' and ``PF'' are applied for the R23 (top) and R23\_lit (bottom) sample. In the top panel, the denominator is the number of targets that satisfy both the fixed criteria (see $\S$\ref{sec:R23_fixed_criteria}) and have AS and PF values above the given threshold. The numerator is the number of these targets that satisfy the period matching criterion. Since the values for the denominator are the same for the R23\_lit sample we provide only the value of the numerator, and the brackets represent the fractional difference between R23 and R23\_lit in a given AS/PF cell.}
\label{tab:match_criteria}

\end{table*}

\subsubsection{Results for individual groups in R23}\label{sec:R23_individual}

\begin{figure}
    \centering
    \includegraphics[width=0.48\textwidth]{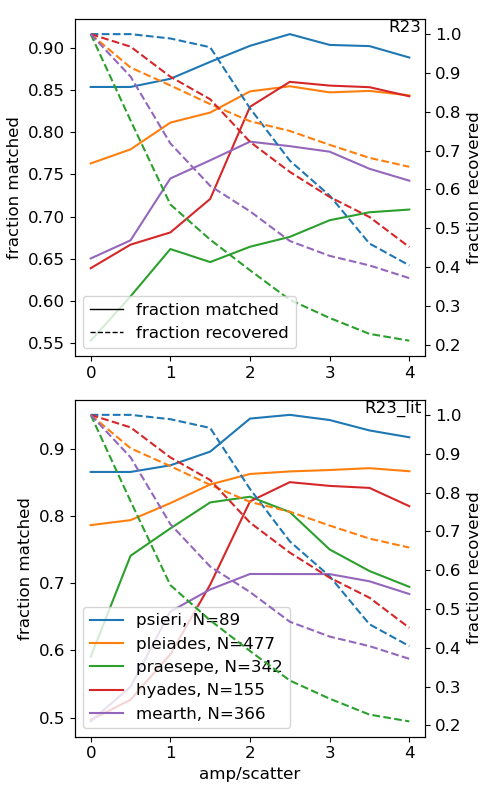}
    \caption{The relationships between both the period matching fraction ($f_{\rm grp}$, solid lines) and fraction of recovered targets ($f_{\rm rec}$, dashed lines) as a function of AS. The top and bottom panels represent the results for the R23 and R23\_lit samples, respectively. The numbers in the legend represent the total number of targets in each group when AS=0.}
    \label{fig:ampscat_matches}
\end{figure}

Our next step is to see how these results look for individual groups. To do so, we apply the fixed criteria, and vary AS in the same way as we did in $\S$\ref{sec:R23_variable_criteria}. Figure~\ref{fig:ampscat_matches}~shows how the fraction of matched targets ($f_{\rm grp}$) and the fraction of recovered targets ($f_{\rm rec}$) changes in each group as a function of AS. As expected, $f_{\rm rec}$ decreases monotonically with increasing AS.

In both samples, the two youngest groups, Psi-Eri (100\,Myr) and Pleiades (125\,Myr), have the highest $f_{\rm grp}$ yield for any value of AS. For Pis-Eri, $f_{\rm grp}=0.86-0.88$ at AS=0. This increases gradually to $\sim0.90$ at AS=4, however, at this stage $f_{\rm rec}$ is only 40 per cent. Therefore, a high yield of $f_{\rm tot}$ and $f_{\rm rec}$ can be recovered at a relatively low value for AS ($\lesssim1.5$) before $f_{\rm rec}$ sharply declines. Similar results are found for the Pleiades, where in both samples there is a steady increase in $f_{\rm grp}$ from $\sim 0.76-0.78$ to $\sim 0.85$ between AS=0 and AS=4. The decline in $f_{\rm rec}$ is less sharp than the Psi-Eri case, dropping to $\sim 0.65$ at AS=4.

The two intermediate-aged groups, Hyades and Praesepe (both $\sim 700\,$Myr) have mixed results. For the Hyades, we find that $f_{\rm grp}$ starts at a value of $\sim 0.65$ and $\sim 0.5$ in the R23 and R23\_lit sample, respectively. They both sharply rise to $f_{\rm grp}=0.82-0.84$ at AS=2, and remain within $\sim 0.03$ of this value between AS=2 and AS=4. The value of $f_{\rm rec}$ falls in a linear fashion down to $\sim 0.45$. In the case of Praesepe, in both samples $f_{\rm grp}$ begins around $0.55$. For the R23 sample, $f_{\rm grp}$ gradually increases to $0.7$ at AS=4. For the R23\_lit sample, $f_{\rm grp}$ increases to a maximum value of $\sim 0.8$ at AS=2. However, this is followed by a gradual {\it decrease} to $\sim 0.7$ at AS=4. We note that $f_{\rm rec}$ falls very sharply, eventually to around $0.2$ (the lowest of all the groups). However, we do not think the decrease in $f_{\rm grp}$ is related to low-number statistics, as there are $\sim 70-90$ targets between $3<$AS$<4$, which is similar to the number of Hyades targets in this AS range. Rather, we identify around a dozen cases where the R23\_lit $P_{\rm rot}$ is double the \TESSILATOR~$P_{\rm fin}$, and 3 cases where this is $>5$ times larger. 

The M-Earth targets in the R23 and R23\_lit samples show similar results in the way $f_{\rm grp}$ changes with AS, but  the R23\_lit are typically $0.05-0.15$ lower. The R23 and R23\_lit sample start at $0.65$ and $0.5$, respectively, both rising to a maximum value around AS=2, where $f_{\rm grp}$ is $0.77$ and $0.7$. These values both slightly decrease by $\sim 0.03$ by AS=4. The $f_{\rm rec}$ values gradually decrease down to $\sim 0.6$ and $0.55$, respectively. It is possible that the R23 sample is in slightly better agreement with \TESSILATOR~because both use TESS data, whereas the R23\_lit values use data with much longer observing baseline. 

\begin{figure*}
    \centering
    \includegraphics[width=0.9\textwidth]{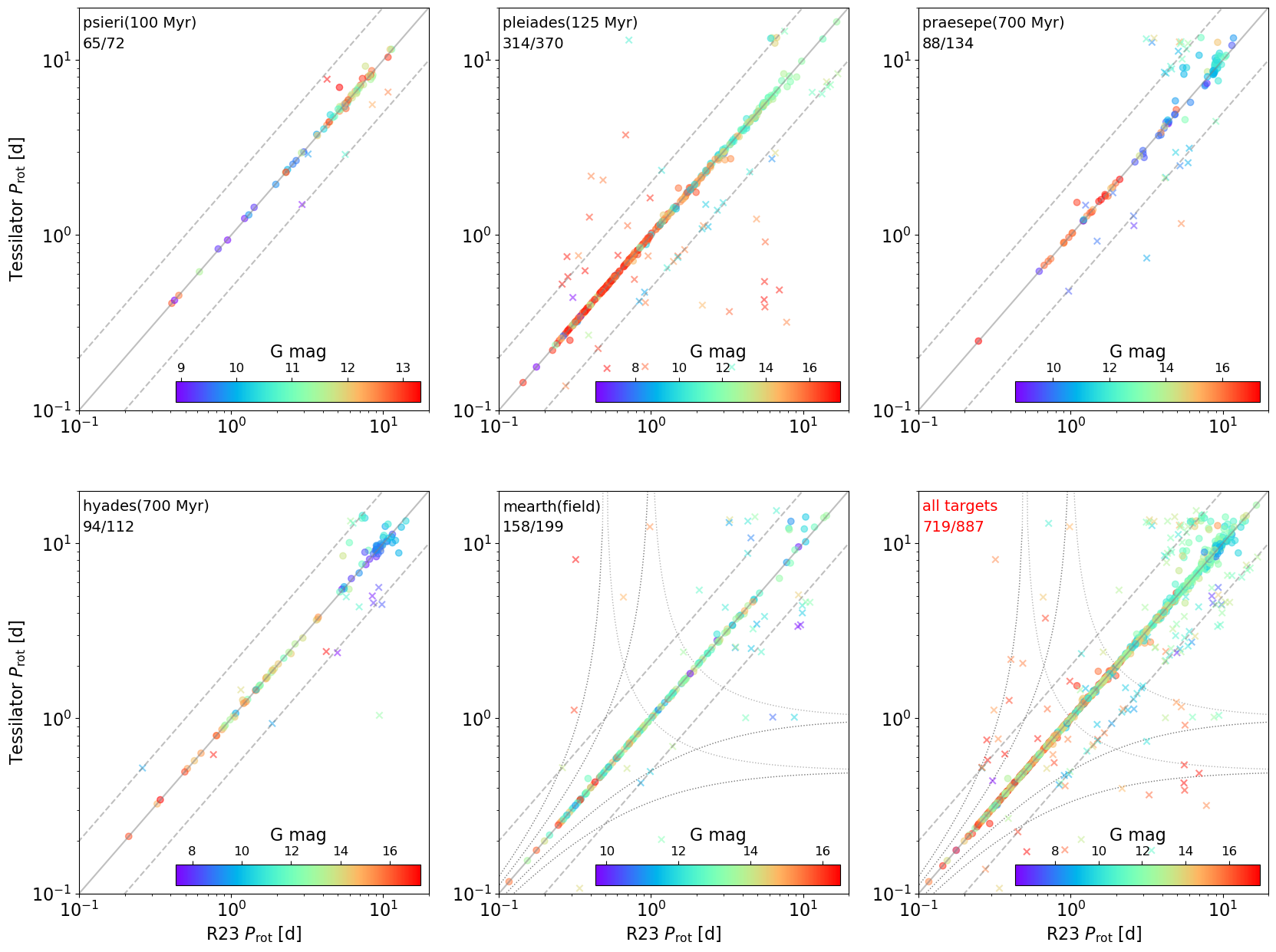} \\
    \caption{A comparison of $P_{\rm rot}$ values between \TESSILATOR~and R23 for the 5 separate groups and the combined sample (bottom-right), in the case where the fixed criteria are applied and AS=2. Filled circles and crosses are the targets that either satisfy, or do not satisfy the period matching criteria, respectively. The numerator and denominator in the fractions shown in the top-left of each panel represent the number of period matches and the number of targets available for testing. Symbols are colour-coded by their Gaia $G$ magnitude and grey lines represent cases where there is equivalence, 1:2, and 2:1. The curved, dashed lines represent the 1-day alias failure modes, using equation 47 in \protect\cite{2018a_VanderPlas}, with $n=0,\pm1$ and $\pm2$.}
    \label{fig:prot_comparison_R23}
\end{figure*}

\begin{figure*}
    \centering
    \includegraphics[width=0.9\textwidth]{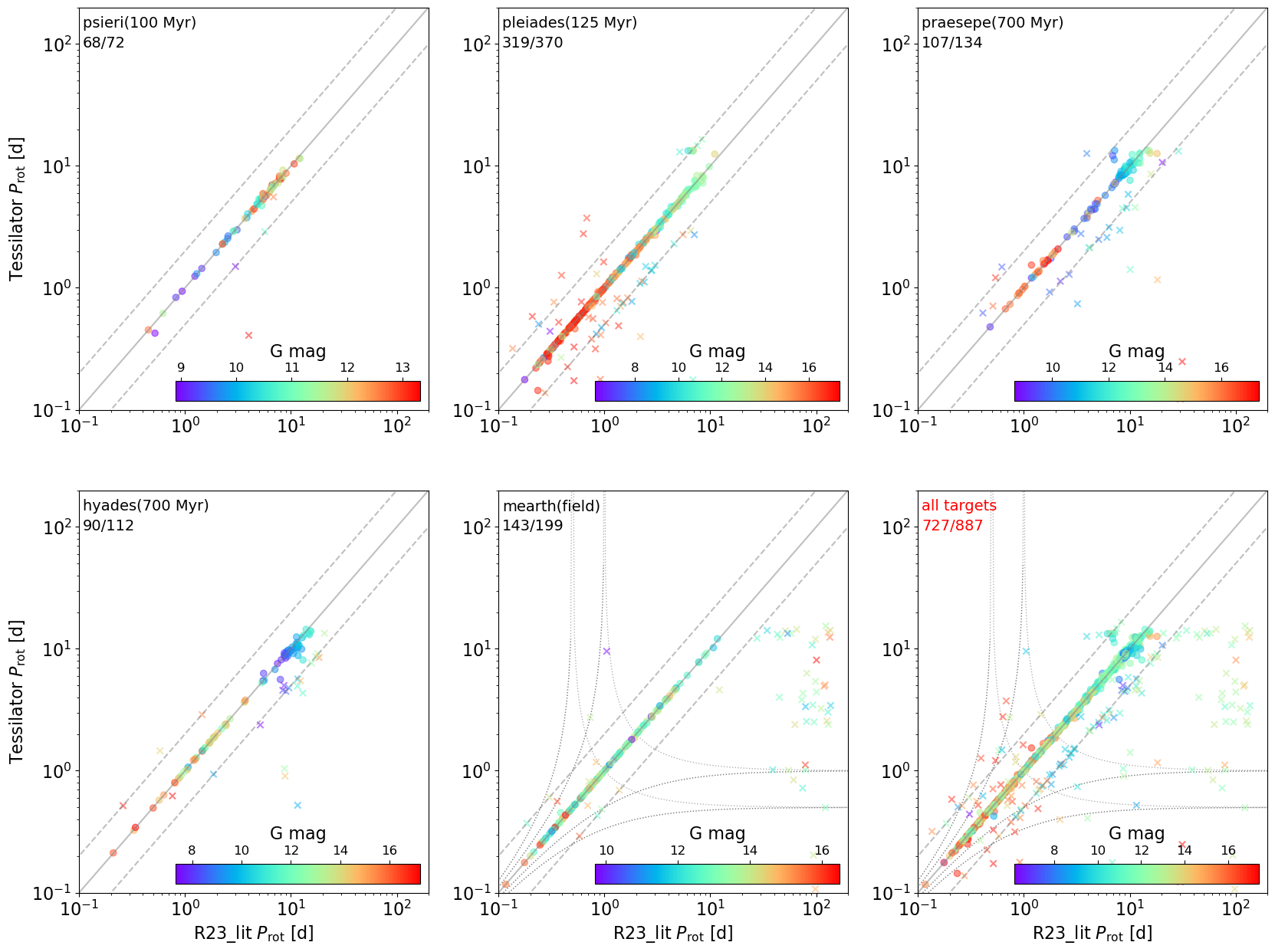} \\
    \caption{The same as Figure~\ref{fig:prot_comparison_R23}, this time with the R23\_lit sample for comparison.}
    \label{fig:prot_comparison_R23_lit}
\end{figure*}

Finally, we present the $P_{\rm fin}$ comparisons for \TESSILATOR~versus R23 and \TESSILATOR~versus R23\_lit in Figures~\ref{fig:prot_comparison_R23}~and \ref{fig:prot_comparison_R23_lit}, respectively. Based on the results of Figure~\ref{fig:ampscat_matches}, we select AP=2.0, which is typically around the maximum yield of $f_{\rm grp}$ for groups in each sample, and ensures the $f_{\rm rec}$ remains relatively large. We also find that the period matches are not correlated in any particular way with Gaia DR3 $G$ magnitude.

The results of the period comparison are provided in Table~\ref{tab:R23_final_periods}. The $P_{\rm fin}$ column contains two sources of uncertainty: firstly, the typical measurement error from all the sectors used to construct $P_{\rm rot}$, and secondly, the standard deviation in $P_{\rm sec}$ (after modification from accounting for half-matches). In this table we also provide a value labelled $N_{\rm half}$, which provides the number of $P_{\rm sec}$ values that were subsequently doubled following the procedure to obtain $P_{\rm fin}$ (see step 2 in $\S$\ref{sec:final_period}).

\begin{table*}
\begin{tabular}{lllllllllll}
\hline\hline
Group & TIC & $G$ & $G_{\rm BP}$ & $G_{\rm RP}$ & $P_{\rm rot,lit}$ & $P_{\rm rot,R23}$ & $P_{\rm rot,fin}$ & $N_{\rm half}$ & $N_{\rm pass}$ & $N_{\rm sec}$ \\
\hline
psieri   & 365332374 & 12.040 & 12.541 & 11.363 & 4.4    & 4.444  & $4.428 \pm 0.262 \pm 0.038$ & 0 & 5 & 5 \\
hyades   & 242985065 & 15.084 & 16.984 & 13.784 & 0.87   & 0.869  & $0.864 \pm 0.010 \pm 0.006$ & 0 & 2 & 2 \\
hyades   & 242944633 & 9.769  & 10.342 & 9.056  & 11.66  & 12.658 & $8.817 \pm 1.424 \pm 1.887$ & 0 & 2 & 3 \\
hyades   & 242991771 & 13.339 & 15.029 & 12.044 & 2.66   & 2.632  & $2.664 \pm 0.099 \pm 0.008$ & 0 & 2 & 2 \\
hyades   & 242946125 & 8.745  & 9.195  & 8.134  & 9.82   & 9.709  & $9.793 \pm 1.434 \pm 0.142$ & 0 & 3 & 3 \\
hyades   & 456863710 & 9.211  & 9.664  & 8.584  & 11.59  & 0.261  & $0.521 \pm 0.002 \pm 0.001$ & 0 & 5 & 5 \\
pleiades & 67785558  & 11.569 & 12.079 & 10.886 & 0.5332 & 0.533  & $0.535 \pm 0.003$           &   & 1 & 1 \\
hyades   & 311151814 & 7.706  & 7.999  & 7.246  & 7.47   & 7.692  & $7.567 \pm 0.760$           &   & 1 & 1 \\
hyades   & 348664634 & 9.604  & 10.135 & 8.924  & 11.6   & 5.525  & $5.653 \pm 0.465 \pm 0.158$ & 0 & 2 & 2 \\
hyades   & 435883224 & 8.108  & 8.502  & 7.542  & 5.52   & 6.173  & $6.291 \pm 0.610$           &   & 1 & 1 \\
\hline
\end{tabular}
\caption{The $P_{\rm rot}$ measurements from our \TESSILATOR~analysis compared with those published in R23, for targets that satisify the fixed and variable criteria in $\S$\ref{sec:R23_fixed_criteria}~and $\S$\ref{sec:R23_variable_criteria}. Columns 1 and 2 provide the name of the individual group and the numeric part of the source identifier in the TESS Input Catalogue (TIC). Columns 3, 4 and 5 represent the Gaia DR3 $G$, $G_{\rm BP}$ and $G_{\rm RP}$ photometric magnitudes, respectively. The $P_{\rm rot}$ values from the literature source used as comparison in R23, and the measurements published by R23 are given in columns 6 and 7, respectively. Column 8 represents the final $P_{\rm rot}$ value derived by \TESSILATOR, where the two uncertainty components comprise firstly of the typical measurement error from each individual sector used and secondly of the standard deviation in the $P_{\rm rot}$ values. Finally columns 9, 10 and 11 are the number of lightcurves that appear to have half-values of $P_{\rm rot}$, the number of sectors used to construct $P_{\rm rot,fin}$ and the number of sectors with $P_{\rm rot}$ measurements prior to selecting $P_{\rm fin}$, respectively. Only the first 10 entries are given here, whereby the full sample of 887 targets is made available in electronic format.}
\label{tab:R23_final_periods}
\end{table*}

Interestingly, comparing \TESSILATOR~with the R23 and R23\_lit sample, we find broadly similar $f_{\rm grp}$ values for Psi-Eri, Pleiades and the Hyades, but for Praesepe, $f_{\rm grp}$ is 0.83 in the R23\_lit sample, compared to 0.66 in R23, which are significantly different fractions at a confidence level $>99$ per cent. There appear to be $\sim 25$ targets in the Praesepe R23 comparison that are mismatches close to either the 1:2 or 2:1 lines. It may be that we would obtain similar $f_{\rm grp}$ values if these were on the equivalence line, however, we also count around a dozen similar targets in the R23\_lit comparison. Therefore, we suspect this discrepancy lies in the methods by which $P_{\rm fin}$ is calculated. Finally, for the M-Earth sample we note that $f_{\rm grp}$ is slightly larger in R23 than R23\_lit (157/199 versus 142/199, marginally significant with a 90 per cent confidence interval). Figure~\ref{fig:prot_comparison_R23_lit}~shows $\sim 40$ M-Earth targets with R23\_lit $P_{\rm rot}$ between 10 and 150\,d. Since M-Earth is a ground-based survey, as a final check, we overlay some of the 1-d alias failure models using equation 47 in \cite{2018a_VanderPlas}, finding very few targets that lie close to these harmonic aliases. Since the observing baseline in the M-Earth mission is several hundreds of days, we believe these to be the true period, which indicates that TESS is incapable measuring a correct $P_{\rm rot}$ in this range.

\subsection{Large field-star $P_{\rm rot}$ surveys}\label{sec:field_star_comparison}

As described in the introduction, the past few years have witnessed an exponential growth in the number of targets with $P_{\rm rot}$ measurements, with the vast majority coming from wide-field variability surveys that target mostly solar-type and low-mass field stars. Since the $P_{\rm rot}$ of a main-sequence low-mass star is usually at least 10\,d \citep{2019a_Amard}, surveys such as {\it Kepler} or the Zwicky Transient Facility (ZTF), with typically year-long observing windows, are very well suited to measuring $P_{\rm rot}$ for these targets. On the other hand, TESS sectors span no more than 27 days and periods can only be reliably measured up to about half of a TESS sector. Whilst efforts are ongoing to ``stitch'' sectors \citep{2022a_Hattori,2024a_Claytor}, time-variable instrument degradation and systematic offsets has revealed this to be a highly problematic, if not impossible task.

Nevertheless, it remains a pertinent exercise to test the \TESSILATOR~using several of the recently published field-star $P_{\rm rot}$ surveys. This can help inform us about the results one could expect when \TESSILATOR~fails because either a target has a $P_{\rm rot}$ too long to be detected by TESS, or the sensitivity-limits and/or contamination makes a TESS analysis unfeasible. We have randomly selected 1000 stars each from 10 catalogues, which represent a mixture of observing facilities and stellar properties (spectral-type, brightness, variability). A brief summary of these surveys is presented in Table~\ref{tab:field_stars}. For simplification, we refer to each survey using the shortened form, provided in column 1. 

\begin{table}
\centering
\begin{tabular}{llll}
\hline\hline
Survey & Instrument & $P_{\rm rot}$ & $G$ \\
\hline
\citealt{2013a_McQuillan}, M13 & Kepler & 0.3-50   & 9.5-16.2 \\
\citealt{2016a_Newton}, N16    & MEarth & 0.1-100+ & 7.2-16.9 \\
\citealt{2018a_Angus}, A18     & Kepler & 1-80     & 8.1-19.3 \\
\citealt{2018a_Oelkers}, O18   & TESS   & 0.5-50   & 4.8-12.9 \\
\citealt{2020a_Reinhold}, R20  & TESS   & 1.2-30   & 11.2-17.3 \\
\citealt{2019a_Santos}, S19    & Kepler & 0.4-100+ & 9.7-17.7 \\
\citealt{2021a_Santos}, S21    & Kepler & 0.4-100+ & 9.7-17.7 \\
\citealt{2022a_Holcomb}, H22   & TESS   & 0.3-20   & 5.3-16.2 \\
\citealt{2022a_Lu}, L22        & ZTF    & 1-100+   & 13.2-17.9 \\
\citealt{2024a_Colman}, C24    & TESS   & 0.5-10   & 5.2-16.1 \\
\hline
\end{tabular}
\caption{Basic properties of the 10 field-star surveys selected for analysis with \TESSILATOR. Columns 3 and 4 denote the approximate literature $P_{\rm rot}$ and $G$-band magnitude range from the 1000 randomly selected targets in each survey.}
\label{tab:field_stars}
\end{table}

We run the \TESSILATOR~using the target identifiers given in the literature source. This is to avoid any ambiguity in coordinate matches, and ensures we measure $P_{\rm rot}$ for the correct target. The \TESSILATOR~is run using all the default parameters (without considerations for CBV corrections). The same selection criteria used to obtain final $P_{\rm rot}$ values from \TESSILATOR~in $\S$\ref{sec:R23_comparison}~are used, as are the conditions to satisfy a matching $P_{\rm rot}$ with the literature value. 

\begin{figure*}
    \centering
    \includegraphics[width=0.9\textwidth]{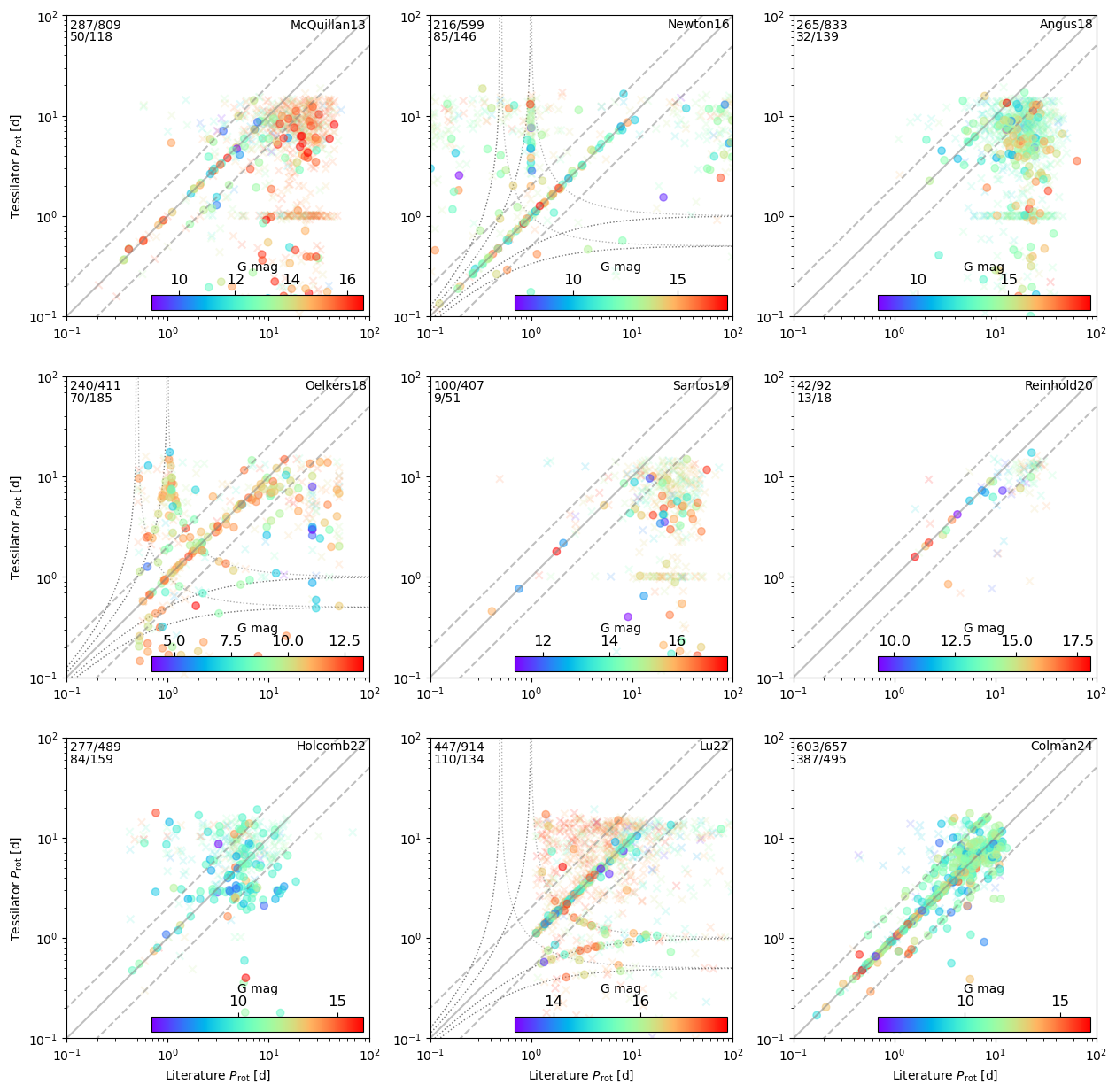} \\
    \caption{Comparison of $P_{\rm rot}$ between \TESSILATOR~and 10 field-star surveys. Crosses and circles indicate targets that fail and pass the quality criteria (given in $\S$\ref{sec:R23_comparison}), respectively, where the fractions in the top-left corners of each panel indicate the ratio of $P_{\rm rot}$ values that are in agreement. Targets are colour-coded by the range of $G$ magnitudes across both the fail/pass sub-samples. Note that we combine the S19 and S21 sample into one panel.}
    \label{fig:field_star_surveys}
\end{figure*}

The results of the $P_{\rm rot}$ comparison are presented in Figure~\ref{fig:field_star_surveys}. There are some visually apparent features in the plot that merit comment. Firstly, it is clear that the fraction of matching $P_{\rm rot}$ values decreases with longer $P_{\rm rot}$, and because of the limit imposed in our initial selection criteria to measure a $P_{\rm rot}$ for each TESS sector, there are no matches when the literature $P_{\rm rot}$ is longer than 20\,d. Although \TESSILATOR~analyses 1000 targets in each survey, there are usually far fewer that end up having a \TESSILATOR~$P_{\rm rot}$ measurement (even before applying criteria), because there are many noisy, flat lightcurves where \TESSILATOR~predicts a long, but unreliable $P_{\rm rot}$. In general, the percentage of matches ($f_{\rm match}$) significantly improves when the criteria are applied, however, with the exception of C24, this comes at the expense of losing more than 3 quarters of the initial matches.

The panels with the highest $f_{\rm match}$ (with criteria applied) tend to be those where the literature source also uses TESS data (e.g., $f_{\rm match}>0.5$ in R20, H22 and C24). The notable exception is O18, where $f_{\rm match}$ is only 38 per cent. We note that many of these targets seem to fall close to the 1-day alias failure modes \citep{2018a_VanderPlas}, and suggest that, because scattered diurnal light is more of an issue in the first few TESS sectors \citep{2018a_Vanderspek}, this may go some way to resolving the discrepancy seen here, where there are approximately 50 targets close to the 1-d alias curves. The comparison with Kepler surveys shows that $f_{\rm match}$ is typically well below half. It is possible that the slightly better agreement with M13 ($f_{\rm match}=0.42$) compared to A18 and S19/S21 ($\sim 0.2-0.25$) is simply that the former sample comprises of Kepler objects of interest that, on average, rotate slower than the full Kepler field sample in M13, and the latter sample was pre-vetted to contain main-sequence targets (which are more likely to have longer $P_{\rm rot}$). Finally, the two ground-based surveys, N16 (MEarth) and L22 (ZTF) have better matches, with $f_{\rm match}$ values of 0.58 and 0.82, respectively. Interestingly, whilst there appear to be no targets with 1-day aliases in N16, a notable number have literature $P_{\rm rot}$ close to 1-day. On the other hand, the L22 panel indicates several dozens of targets that match closely to the 1-day alias harmonics. The good agreement in both of these might be because both surveys target early M-stars, that typically take longer to spin-down, but this may also be indicative of an effective approach to screen out poor quality data from \TESSILATOR~lightcurves. 

\section{Conclusions}\label{sec:conclusions}

We present our new \TESSILATOR~code, which is available online. Whilst other software tools exist for working with TESS data, \TESSILATOR~is unique because it provides an all-in-one package for downloading image files, extracting aperture photometry, assessing the background contamination, processing lightcurves, performing LSP analyses and calculating $P_{\rm rot}$. The code requires minimal input from the user and can be executed with simple command-line prompts. In this paper, we describe all our methods in a conceptual way, providing examples from real data to guide the narrative. We tested our code by comparing our \TESSILATOR~periods for a sample of 1560 stars comprising of members of 4 young groups and a field-star sample, with periods measured from two independent sources. The results indicate very good (65-90 per cent) agreement across all groups, which validates \TESSILATOR~as a reliable method for measuring $P_{\rm rot}$. We also compare our software with several large field star surveys, finding reasonable agreement for TESS-based samples, as well as  survey with longer observing baselines, when the $P_{\rm rot}$ is within the observing capabilities of TESS. Unsurprisingly, for targets with $P_{\rm rot}>13\,$d, \TESSILATOR~is generally unable to find a matching value.

With the period validation successfully implemented, we believe that \TESSILATOR~can now be used for a large number of scientific applications. To name a few, these include: understanding the evolution of $P_{\rm rot}$ and angular momentum by observing targets in young open clusters; testing new empirical models of Gyrochronology; investigating the connection between age, rotation and activity (or other parameters that vary with age/rotation); and identifying fast rotators as potentially young stars in large-scale surveys. Regarding this final application, \TESSILATOR~has already been used as part of the target selection strategy for a 5-year spectroscopic observing campaign to identify nearby, young stars using the 4-metre Multi-Object Spectroscopic Telescope (MOST). From an initial sample of $\sim 1.1\times10^{6}$ G/K-type stars within 500\,pc, \TESSILATOR~identified $\sim 20,\,000$ fast rotators (and therefore young star candidates) which are scheduled for observation as part of the 4MOST Survey of Young Stars (4SYS).

Future \TESSILATOR~releases are anticipated, which will be largely driven by feedback from users. In our next version of the code, we aim to implement data-driven methods to improve reliability of measurements, which will lead to better ways to quantify the quality indicators. We hope the community find the tool a useful, unique addition to the large suite of software already available for use, and we are planning to use \TESSILATOR~for several scientific applications in the near future.

\section*{Acknowledgements}

We are indebted to an anonymous referee, who provided constructive and thoughtful comments that led to a significant improvement in the manuscript. A.S. Binks acknowledges financial support from the Eberhard Karls Universit\"at T\"ubingen and from the Massachusetts Institute of Technology. H.M.\ G\"unther was supported by the National Aeronautics and Space Administration through Chandra Award Number AR3-24001X issued by the Chandra X-ray Center, which is operated by the Smithsonian Astrophysical Observatory for and on behalf of the National Aeronautics Space Administration under contract NAS8-03060. This paper includes data collected from the TESS mission, obtained from the MAST data archive at the Space Telescope Science Institute (STScI). This work has made use of data from the European Space Agency (ESA) mission Gaia (\url{https://www.cosmos.esa.int/gaia}), processed by the Gaia Data Processing and Analysis Consortium (DPAC, \url{https://www.cosmos.esa.int/web/gaia/dpac/consortium}). Funding for the DPAC has been provided by national institutions, in particular the institutions participating in the Gaia Multilateral Agreement. The authors acknowledge support by the High Performance and Cloud Computing Group at the Zentrum für Datenverarbeitung of the University of Tübingen, the state of Baden-Württemberg through bwHPC and the German Research Foundation (DFG) through grant no INST 37/935-1 FUGG.

\section*{Data Availability}

\TESSILATOR~is a publicly-available code, which is currently hosted at the following Git-Hub link: \url{https://github.com/alexbinks/tessilator}. The code operates under an MIT-style license. The full version of Table~\ref{tab:R23_final_periods}~is available in electronic format, and we plan to make the data accessible with the VizieR catalog service, provided by Centre de donn\'ees astronomiques de Strasbourg. The data used to construct Figure~\ref{fig:field_star_surveys}~is available upon request.



\bibliographystyle{mnras}
\bibliography{bibliography} 




\appendix

\section{CBV tests used for selecting lightcurves}\label{sec:CBV_tests}

In $\S$\ref{sec:CBV_corrections}~we describe that there are two separate tests (CBV test 1 and CBV test 2) used to determine whether a CBV-corrected lightcurve should be selected from which to calculate $P_{\rm rot}$. Each test uses a scoring system, where the CBV-corrected lightcurve must score higher than the original lightcurve (a draw results in the original lightcurve being selected). For CBV test 1, the fluxes from each lightcurve are first normalised using the median flux value ($f_{\rm med}$). We shall label these normalised fluxes as $f_{\rm norm}$. For CBV test 1, we have the following rounds:

\begin{itemize}
\item [(1a)] {\bf Which lightcurve has the least extreme outliers?}

The normal-adjusted MAD flux value from $f_{\rm norm}$ is calculated (labelled $f_{\rm MAD}$), and the lightcurve with the least number of $f_{\rm norm}$ values outside the range $f_{\rm med} \pm f_{\rm MAD}$ is awarded the point.

\vspace{0.3cm}
\item [(1b)] {\bf Which lightcurve has the lowest $f_{\rm MAD}$ value?}

The lightcurve with the lowest $f_{\rm MAD}$ value wins the point.

\vspace{0.3cm}
\item [(1c)] {\bf Which lightcurve is best matched to a sinusoidal fit?}

A sinusoidal fit of the form $f_{\rm norm} = f_{0} + A\sin\left(\frac{2\pi}{P_{\rm rot}}t + \phi \right)$~is used, where $f_{0}$ is a constant, $A$ is the amplitude, $P_{\rm rot}$ is the period, $\phi$ is the phase and $t$ is the time coordinate to be fitted. The lightcurve that provides the lowest reduced $\chi^{2}$ score wins the point.
\end{itemize}

\vspace{0.6cm}

For CBV test 2, we use features from the periodogram analysis to assess which lightcurve gives the best results. The lightcurves used here have been passed through all the processing steps. There are 5 parts to this test, which are as follows:

\begin{itemize}
\item [(2a)] {\bf Straight line or sinusoidal fit?}

Using the AIC procedure described in $\S$\ref{sec:detrending}, determine whether the best fit corresponds to a straight line or a sinusoid. If the latter, then a point is scored.

\vspace{0.3cm}
\item [(2b)] {\bf How smooth are the lightcurves?}

If the ``jump flag'' (described in $\S$\ref{sec:outputs_and_finalprot}) returns a value of 1, a point is scored.

\vspace{0.3cm}
\item [(2c)] {\bf False Alarm Probabilities}

Measure the ratio between the highest LSP power output and False Alarm Probability corresponding to a 1\,per cent threshold. Whichever lightcurve has the highest value wins the point.

\vspace{0.3cm}
\item [(2d)] {\bf The amplitude-to-scatter ratio}

The ratio of the amplitude and the typical scatter of the phase-folded lightcurve (described in $\S$\ref{sec:phase_folding}) is calculated for both lightcurves. The highest value wins the point.

\vspace{0.3cm}
\item [(2e)] {\bf Total number of data points}

The lightcurve with the largest number of data points wins the point.
\end{itemize}

\bsp	
\label{lastpage}
\end{document}